\documentstyle[12pt]{article}

\global\arraycolsep=2pt
\makeatletter
\def\fmslash{\@ifnextchar[{\fmsl@sh}{\fmsl@sh[0mu]}}
\def\fmsl@sh[#1]#2{%
  \mathchoice
    {\@fmsl@sh\displaystyle{#1}{#2}}%
    {\@fmsl@sh\mboxstyle{#1}{#2}}%
    {\@fmsl@sh\scriptstyle{#1}{#2}}%
    {\@fmsl@sh\scriptscriptstyle{#1}{#2}}}
\def\@fmsl@sh#1#2#3{\m@th\ooalign{$\hfil#1\mkern#2/\hfil$\crcr$#1#3$}}
\makeatother

\begin{document}

\vspace{1.cm}
\begin{center}
\Large\bf Decay Widths of $B_1$ and $B_2^\ast$ up to the order
${\cal O}(1/m_Q)$ in HQET
\end{center}
\vspace{0.5cm}
\begin{center}
{ Yuan-Ben Dai and Shi-Lin Zhu}\\\vspace{3mm}
{\it Institute of Theoretical Physics,
 Academia Sinica, P.O.Box 2735, Beijing 100080, China }
\end{center}

\vspace{1.5cm}
\begin{abstract}
We first improve the previous results on the masses of the doublets
$(0^+,1^+)$ and $(1^+,2^+)$ with QCD sum rules up 
to the order of ${\cal O}(1/m_Q)$ in HQET. 
With the mass matrix of the two $1^+$ states,
we obtain the mixing angle for the two states and construct the
interpolating currents which couple to only one of the two eigen states
up to the order ${\cal O}(1/m_Q)$. Using these results we derive the
light-cone sum rules for the amplitutes of the pionic decay of $B_1$
and $B_2^*$ up to the order ${\cal O}(1/m_Q)$ in HQET. The ${\cal
O}(1/m_Q)$ terms in these amplitutes are found to be small. When applied to 
$D_1$ and $D_2^*$ the theoretical results are in good agreement with
experimental data though the theoretical uncertainty is sizable for the
$D$ system.

\end{abstract}


\pagenumbering{arabic}
%%%%%%%%%%%%%%%%%%%%%%%%%%%%%%%%%%%%%%%%%%%%%%%%%%%%%%%%%%%%%%%%%%%%%%%%%%%%%%%%%%%%%
%%%%%%%%%%%%%%%%%%%%%%%%%%%%%%%%%%%%%%%%%%%%%%%%%%%%%%%%%%%%%%%%%%%%%%%%%%%%%%%%%%%%%
\section{ Introduction}
\label{sec1} 
Important progress has been achieved in understanding the properties of
heavy mesons composed of a heavy quark and a light quark with the development 
of the heavy quark effective theory (HQET) \cite{grinstein}. HQET provides a 
systematic expansion of the heavy hadron spectra and transition amplitude
in terms of $1/m_Q$, where $m_Q$ is the heavy quark mass.
Of course one has to employ some specific 
nonperturbative methods to arrive at the detailed predictions. Among the
various nonpeturbative methods, QCD sum rules is useful to extract the 
low-lying hadron properties \cite{svz}. The propertes of the ground state
heavy mesons have been studied with the QCD sum rules in HQET in \cite{bagen},
\cite{ball}. In \cite{huang} the mass of the lowest excited heavy meson 
doublets $(2^+,1^+)$ and $(1^+,0^+)$ were studied with QCD sum rules in the
heavy quark effective theory (HQET) up to the order of ${\cal O}(1/m_Q)$. 
The widths for pionic decays of the
lowest two excited doublets $(2^+,1^+)$ and $(1^+,0^+)$ is calculated 
with conventional QCD sum rules in the leading order of HQET in \cite{ybd97}.
Properties of excited heavy mesons were investigated also with QCD sum
rules in full QCD in \cite{col91}. Deacy of $(1^+,0^+)$ were
investigated with light-cone sum rules \cite{bely95} in \cite{col97}. 
The pionic couplings between the lowest three doublets $(2^+,1^+)$, 
$(1^+,0^+)$ and $(1^-,0^-)$ of heavy mesons are investigated in HQET 
using the light-cone sum rules \cite{zhu1}. In \cite{zhu2} the pionic
couplings of the heavy baryons are calculated with the same technique.

One difficult problem encountered in studying the decay widths of excited heavy mesons
with QCD sum rules is the following. Except for the lowest states $0^-$, 
$1^- $, the spectra contains a pair of states for any spin-parity
$J^P$. An interpolating current with definite spin-parity in general
couples with both two states.This parity is especially serious for the
two $1^+$ states.All theoretical calculations indicate that they are
close in mass values but quite different in magnitude of their decay widths.
One of the two $1^+$ states is a narrow resonance
decaying mainly by emitting a $D$ wave pion, while the other one is a very
wide resonance decaying mainly by emitting a ${\cal S}$ wave pion. An interpolating
current used for the narrow $1^+$ state with a small coupling to the other $%
1^+$ state may cause sizable error in the result of calculation. Only
in the $m_Q\to\infty$ limit is there a conserved quantum number $j_{\ell}$,
the angular momentum of the light component, which can be used to
differentiate the two states. Therefore, HQET has the important and unique 
advantage for this purpose. Corrections for finite $m_Q$ can be calculated
order by order in HQET.The interpolating currents which couple to only
one of the two $J^P$ states have been given in \cite{huang} and used in
\cite{ybd97}.

In this work we shall calculate the decay widths of the doublet $(1^+, 2^+)$
to the first order in $1/m_Q$ using the light-cone sum rule approach.
As the necessary steps we present the results from sum rules in HQET for masses
of these excited states and their coupling constants to the
interpolating currents at the leading and next-to leading order in
section 2 and section 3 respectivly. A large part of contents of these sections
have been obtained in \cite{huang}.The new things include the addition
of the leading ${\cal O}(\alpha_s)$ perturbation term for the chromo-magnetic
splitting $\Sigma$ and the results for corrections to coupling constants at the order
${\cal O}(1/m_Q)$. We also improve the numerical analysis of the sum rules at
the leading order. The continuum threshold $\omega_c$ 
and the working region for the Borel parameter $T$ 
determined from the stability criterum of the sum rules in the leading order  			
are used in the sum rules for $\Sigma$ term and the kinetic energy term ${\cal K}$ 
in the corrections to the masses. Solving the mass matrix of the two distinct $1^+$ states,
we arrive at the mixing angle for these two states. Using this angle the
interpolating currents which couple to only one of the $1^+$ states up
to the order $1/m_Q$ are constructed in section \ref{sec4}. 
In section \ref{sec5} and \ref{sec6} we employ the constructed currents 
and the light-cone QCD sum rules (LCQSR) 
\cite{bely95} in HQET to calculate  
the pionic decay amplitudes of $B_1$ and $B_2^*$ up to the order
${\cal O}(1/m_Q)$ taking into account the
mixing of two $1^+$ states. The results are summarized and discussed 
in section \ref{sec9}.

%%%%%%%%%%%%%%%%%%%%%%%%%%%%%%%%%%%%%%%%%%%%%%%%%%%%%%%%%%%%%%%%%%%%%%%%%%%%%%%%%%%%%%
%%%%%%%%%%%%%%%%%%%%%%%%%%%%%%%%%%%%%%%%%%%%%%%%%%%%%%%%%%%%%%%%%%%%%%%%%%%%%%%%%%%%%%
\section{Two-point QCD sum rules}
\label{sec2}
The proper interpolating current $J_{j,P,j_{\ell}}^{\alpha_1\cdots\alpha_j}$
for the state with the quantum number $j$, $P$, $j_{\ell}$ in HQET was
given in \cite{huang}. These currents were proved to satisfy the following conditions 
\begin{eqnarray}
\label{decay}
\langle 0|J_{j,P,j_{\ell}}^{\alpha_1\cdots\alpha_j}(0)|j',P',j_{\ell}^{'}\rangle&=&
f_{Pj_l}\delta_{jj'}
\delta_{PP'}\delta_{j_{\ell}j_{\ell}^{'}}\eta^{\alpha_1\cdots\alpha_j}\;,\\
\label{corr}
i\:\langle 0|T\left (J_{j,P,j_{\ell}}^{\alpha_1\cdots\alpha_j}(x)J_{j',P',j_{\ell}'}^{\dag
\beta_1\cdots\beta_{j'}}(0)\right )|0\rangle&=&\delta_{jj'}\delta_{PP'}\delta_{j_{\ell}j_{\ell}'}
(-1)^j\:{\cal S}\:g_t^{\alpha_1\beta_1}\cdots g_t^{\alpha_j\beta_j}\nonumber\\[2mm]&&\times\:
\int \,dt\delta(x-vt)\:\Pi_{P,j_{\ell}}(x)
\end{eqnarray}
in the $m_Q\to\infty$ limit, where $\eta^{\alpha_1\cdots\alpha_j}$ is the
polarization tensor for the spin $j$ state, $v$ is the velocity of the heavy
quark, $g_t^{\alpha\beta}=g^{\alpha\beta}-v^{\alpha}v^{\beta}$ is the
transverse metric tensor, ${\cal S}$ denotes symmetrizing the indices and
subtracting the trace terms separately in the sets $(\alpha_1\cdots\alpha_j)$
and $(\beta_1\cdots\beta_{j})$, $f_{P,j_{\ell}}$ and $\Pi_{P,j_{\ell}}$ are
a constant and a function of $x$ respectively which depend only on $P$ and $%
j_{\ell}$. Because of equations (\ref{decay}) and (\ref{corr}), the sum rule
in HQET for decay widths derived from a correlator containing such currents
receives no contribution from the unwanted states with the same spin-parity
as the states under consideration in the $m_Q\to\infty$. Starting from the
calculations in the leading order, the decay amplitudes for finite $m_Q$ can
be calculated unambiguously order by order in the $1/m_Q$ expansion in HQET.

We shall confine our study to the doublets $(0^+,1^+)$ and $(1^+,2^+)$ here.
There are two possible choices for currents creating $0^+$ and $1^+$ of the doublet $(0^+,1^+)$, either
\begin{eqnarray}
\label{curr1}
J^{\dag}_{0,+,2}&=&\frac{1}{\sqrt{2}}\:\bar h_vq\;,\\
\label{curr2}
J^{\dag\alpha}_{1,+,2}&=&\frac{1}{\sqrt{2}}\:\bar h_v\gamma^5\gamma^{\alpha}_tq\;,
\end{eqnarray}
or
\begin{eqnarray}
\label{curr3}
J^{'\dag}_{0,+,2}&=&\frac{1}{\sqrt{2}}\:\bar h_v(-i)\not\!{\cal D}_tq\;,\\
\label{curr4}
J^{'\dag}_{1,+,2}&=&\frac{1}{\sqrt{2}}\:\bar h_v\gamma^5\gamma^{\alpha}_t(-i)\not\!{\cal D}_tq\;.
\end{eqnarray}
Similarly, there are two possible choices for the currents 
creating $1^+$ and $2^+$ of the doublet $(1^+,2^+)$. One is 
\begin{eqnarray}
\label{curr5}
J^{\dag\alpha}_{1,+,1}&=&\sqrt{\frac{3}{4}}\:\bar h_v\gamma^5(-i)\left(
{\cal D}_t^{\alpha}-\frac{1}{3}\gamma_t^{\alpha}\not\!{\cal D}_t\right)q\;,\\
\label{curr6}
J^{\dag\alpha_1,\alpha_2}_{2,+,1}&=&\sqrt{\frac{1}{2}}\:\bar h_v
\frac{(-i)}{2}\left(\gamma_t^{\alpha_1}{\cal D}_t^{\alpha_2}+\gamma_t^{\alpha_2}{\cal D}_t^{\alpha_1}-\frac{2}{3}\;g_t^{\alpha_1\alpha_2}\not\!{\cal D}_t\right)q\;.
\end{eqnarray}
Another choice is obtained by adding a factor $-i\not\!{\cal D}_t$ to (\ref{curr5}) and (\ref{curr6}). Note
that, without the last term in the bracket in (\ref{curr5}) the current
would couple also to the $1^+$ state in the doublet $(0^+,1^+)$ even in
the limit of infinite $m_Q$.

For the doublet $(0^+,1^+)$, when the currents ${J'}_{0,+,1/2}$, ${J'}_{1,+,1/2}$ 
in (\ref{curr3}), (\ref{curr4}) are
used the sum rule (same for the two states) after the Borel transformation is found to be
\begin{eqnarray}
\label{form1}
&&{f'}_{+,1/2}^2e^{-2\bar\Lambda/{T}}=\frac{3}{2^6\pi^2}\int_0^{\omega_c}\omega^4e^{-\omega/{T}}d\omega-\frac{1}{2^4}\,m_0^2\,\langle\bar qq\rangle\;.
\end{eqnarray}
The corresponding formula when the current $J_{0,+,2}$ and $J_{1,+,2}$ in (\ref{curr1}) and (\ref{curr2}) are used instead of
${J'}_{0,+,2}$ and ${J'}_{1,+,2}$ is the following 
\begin{eqnarray}
\label{form2}
&&f_{+,1/2}^2e^{-2\bar\Lambda/{T}}=\frac{3}{16\pi^2}\int_0^{\omega_c}\omega^2e^{-\omega/{T}}d\omega+\frac{1}{2}\,\langle\bar qq\rangle-{1\over 8T^2}\,m_0^2\,\langle\bar qq\rangle\;.
\end{eqnarray}
When the currents (\ref{curr5}) and (\ref{curr6}) are used the sum rule for
the $(1^+,2^+)$  doublet is found to be
\begin{eqnarray}
\label{form3}
&&f_{+,3/2}^2e^{-2\bar\Lambda/{T}}={1\over 2^6\pi^2}\int_0^{\omega_c}\omega^4e^{-\omega/{T}}d\omega
-\frac{1}{12}\:m_0^2\:\langle\bar qq\rangle-{1\over 2^5}\langle{\alpha_s\over\pi}G^2\rangle T\;.
\end{eqnarray}
Here $m_0^2\,\langle\bar qq\rangle=\langle\bar qg\sigma_{\mu\nu}G^{\mu\nu}q\rangle$.
These sum rules have been obtained in \cite{huang}. In the derivations, only
terms of the lowest order 
in perturbation and operators of dimension less than six have been included. 

For the QCD parameters entering the theoretical expressions, we use the following standard values
\begin{eqnarray}
\label{parameter}
\langle\bar qq\rangle&=&-(0.24 ~\mbox{GeV})^3\;,\nonumber\\
\langle\alpha_s GG\rangle&=&0.038 ~\mbox{GeV}^4\;,\nonumber\\
m_0^2&=&0.8 ~\mbox{GeV}^2\;.
\end{eqnarray}
Here we redo the numerical analysis of the sum rules with the following
criteria.The high-order power corrections is less than $30\%$ of the
pertuabation term without the cuttoff $\omega_c$
and the contribution of the pole term is larger than $40\%$ of the continuum
contribution given by the perturbation integeral in the region $\omega
> \omega_c$. $\omega_c$ is constrained to be smaller than 3.2 Gev
which is roughly the expected position of the pole of the radial excited state
with the same quantum numbers. Subjected to these criteria we arrive at the stability
region of the sum rules $\omega_c=2.8-3.2$ GeV, $T=0.43-0.73$ GeV for (\ref{form1}), 
$\omega_c=2.8-3.2$ GeV, $T=0.82-1.1$ GeV for (\ref{form2}), 
and $\omega_c=2.8-3.2$ GeV, $T=0.54-0.73$ GeV for (\ref{form3}). 
The results for $\bar\Lambda$ for three cases are the following
\begin{equation}
\label{result1}
\bar\Lambda({1/2}^+)=1.15\pm 0.10 ~~\mbox{GeV},
\end{equation}
for the doublet $(0^+,1^+)$ when the currents
(\ref{curr1}), (\ref{curr2}) without the derivative is used.
\begin{equation}
\label{result2}
\bar\Lambda({1/2}^+)=0.95\pm 0.10 ~~\mbox{GeV},
\end{equation}
for the same doublet when the currents (\ref{curr3}), (\ref{curr4}) with
the derivative is used.
\begin{equation}
\label{result3}
{\bar \Lambda}({3/2}^+)=0.82\pm 0.10 ~~\mbox{GeV},
\end{equation}
for the doublet $(1^+,2^+)$.
These results shall be used in the following sections.

In the following sections we also need the values of f's.
\begin{equation}
\label{ff1}
f_{+, 3/2}=0.19\pm 0.03 ~~\mbox{GeV}^{5/2},
\end{equation}
\begin{equation}
\label{ff2}
f_{+, 1/2}^{\prime}=0.37\pm 0.06 ~~\mbox{GeV}^{5/2},
\end{equation}
\begin{equation}
\label{ff3}
f_{+, 1/2}=-(0.40\pm 0.06) ~~\mbox{GeV}^{3/2} .
\end{equation}
Note it is impossible to determine the absolute sign of f's from 
QCD sum rule approach. From the equation of motion
we can derive \cite{huang}:
\begin{equation}\label{ddd}
f_{+, 1/2}^{\prime}=-{\bar  \Lambda}_{+,{1\over 2}} f_{+, 1/2}\; .
\end{equation}
In this work we adopt the convention of the positive $f'_{+, 1/2}$, 
$f_{+, 3/2}$.

%%%%%%%%%%%%%%%%%%%%%%%%%%%%%%%%%%%%%%%%%%%%%%%%%%%%%%%%%%%%%%%%%%%%%%%%%%%%%%%%%%%%%
%%%%%%%%%%%%%%%%%%%%%%%%%%%%%%%%%%%%%%%%%%%%%%%%%%%%%%%%%%%%%%%%%%%%%%%%%%%%%%%%%%%%%
\section{The sum rules at the ${\cal O}(1/m_Q)$ order}
\label{sec3}
Inserting the heavy meson eigen-state of the Hamiltonian up to 
the order ${\cal O}(1/m_Q)$ into the correlator
\begin{eqnarray}
\label{del}
i\int d^4xe^{ik\cdot x}\langle 0|T\left(J_{j,P,j_l}^{\alpha_1\cdots\alpha_j}(x)
J_{j,P,j_l}^{\dag\beta_1\cdots\beta_{j}}(0)\right )|0\rangle\;,
\end{eqnarray}
the pole term on the hadron side becomes
\begin{eqnarray}
\label{m-pole}
\Pi(\omega)_{pole}={(f+\delta f)^2\over 2(\bar\Lambda+\delta m)-\omega}={f^2\over 2\bar\Lambda-\omega}-{2\delta mf^2\over (2\bar\Lambda-\omega)^2}+{2f\delta f\over 2\bar\Lambda-\omega}\;,
\end{eqnarray}
where $\delta m$ and $\delta f$ are of the order ${\cal O}(1/m_Q)$.

To extract $\delta m$ in (\ref{m-pole}) 
we follow the approach of \cite{ball} and consider
the three-point correlation functions
\begin{eqnarray}
\label{delta}
\delta_O\Pi^{\alpha_1\cdots\alpha_j,\beta_1\cdots\beta_j}_{j,P,i}(\omega,\omega^{'})
=i^2\int d^4xd^4ye^{ik\cdot x-ik'\cdot y}\langle 0|T\left(J_{j,P,i}^{\alpha_1\cdots\alpha_j}(x)\;O(0)\;J_{j,P,i}^{\dag\beta_1\cdots\beta_{j}}(y)\right )|0\rangle\;,
\end{eqnarray}
where $O={\cal K}$ or ${\cal S}$ is the kinetic energy or the spin
dependent term of the ${\cal O}(1/m_Q)$ lagrangian of the HQET. 
The scalar function corresponding to (\ref{delta}) can be represented as the double dispersion integral
\begin{eqnarray}
\label{delta1}
\delta_O\Pi(\omega,\omega^{'})={1\over\pi^2}\int{{\rho}_o(s,s')dsds'\over(s-\omega)(s'-\omega^{'})}\;.
\end{eqnarray}
The pole parts for them are
\begin{eqnarray}
\label{polek}
&&\delta_{\cal K}\Pi(\omega,\omega^{'})_{pole}=
{f^2K\over(2\bar\Lambda-\omega)(2\bar\Lambda-\omega^{'})}
+{f^2G_{\cal K}(\omega^{'})\over 2\bar\Lambda-\omega}+{f^2G_{\cal K}(\omega)\over 2\bar\Lambda-\omega^{'}}\;,\\[2mm]
\label{poles}
&&\delta_{\cal S}\Pi(\omega,\omega^{'})_{pole}=
{d_Mf^2\Sigma\over(2\bar\Lambda-\omega)(2\bar\Lambda-\omega^{'})}
+d_Mf^2\left[{G_{\cal S}(\omega^{'})\over 2\bar\Lambda-\omega}+{G_{\cal S}(\omega)\over 2\bar\Lambda-\omega^{'}}\right]\;,
\end{eqnarray}
where
\begin{eqnarray}
&&{K_{j,P,j_l}}=\langle j,P,j_l|{\bar h_v\,(i D_\perp)^2 h_v}|j,P,j_l\rangle\;,\\
&&{2d_M\Sigma_{j,P,j_l}}=\langle j,P,j_l|{
   \bar h_v\,g\sigma_{\mu\nu} G^{\mu\nu} h_v}|j,P,j_l\rangle\;,\\
&&d_M=d_{j,j_l},\hspace{2.5mm}d_{j_l-1/2,j_l}=2j_l+2,\hspace{2.5mm}d_{j_l+1/2,j_l}=-2j_l.
\end{eqnarray}
Let $\omega=\omega^{'}$ in (\ref{polek}) and (\ref{poles}), and compare 
it with (\ref{m-pole})
one obtains\cite{ball}
\begin{eqnarray}
\label{delm}
\delta m=-{1\over 4m_Q}(K+d_MC_{mag}\Sigma)\;.
\end{eqnarray}
The simple pole term in (\ref{polek}) and (\ref{poles}) comes from 
the region in which $s(s')=2\bar\Lambda$ and $s(s')$ is at the pole for a 
radial excited state or in the continuum. These terms are suppressed by making double Borel transformation for bo
th $\omega$ and $\omega'$. One obtains thus the sum rules for $K$ and $\Sigma$ as
\begin{eqnarray}
\label{form4}
&&f^2K\,e^{-2\bar\Lambda/{T}}=\int_0^{\omega_c}\int_0^{\omega_c}d\omega d\omega'e^{-(\omega+\omega')/2T}\rho_{\cal K}(\omega,\omega')\;,\\
&&f^2\Sigma\,e^{-2\bar\Lambda/{T}}=\int_0^{\omega_c}\int_0^{\omega_c}d\omega d\omega'e^{-(\omega+\omega')/2T}\rho_{\cal S}(\omega,\omega')\;,
\end{eqnarray}
where the spectral densities are obtained from straightforward calculations in HQET.
													
Confining us to the leading order of perturbation and the operators
with dimension $D\leq 5$ in OPE, we find 
for the $j_l^P=\displaystyle{{1\over 2}^+}$ doublet, 
\begin{eqnarray}
\label{form-k1d}
&&f^2K\,e^{-2\bar\Lambda/{T}}=-{3\over 2^7\pi^2}\int_0^{\omega_c} \omega^6\;e^{-\omega/T}d\omega+{3\over 2^4\pi}\langle\alpha_sGG\rangle\;T^3\;,\\
\label{form-s1d}
&&f^2\Sigma\,e^{-2\bar\Lambda/{T}}=
\int_0^{\omega_c}\int_0^{\omega_c} \rho (s, s^\prime )\;e^{-{s+s^\prime \over 2T}}dsds^\prime
+{1\over 48\pi}\langle\alpha_sGG\rangle \;T^3\;,
\end{eqnarray}
with $\rho (s,s^\prime )={1\over 96}{\alpha_s(2T)\over \pi^3} C_{mag}
ss^\prime \{ {s^\prime}^2 (3s-s^\prime ) \theta (s-s^\prime ) 
+ (s\leftrightarrow s^\prime ) \}$
when the currents (\ref{curr3}) and (\ref{curr4}) are used and
\begin{eqnarray}
\label{form-k1}
&&f^2K\,e^{-2\bar\Lambda/{T}}=-{3\over 2^5\pi^2}\int_0^{\omega_c} \omega^4\;e^{-\omega/T}d\omega-{1\over 2^5\pi}\langle\alpha_sGG\rangle\;T-{3\over 8}\;m_0^2\;\langle\bar qq\rangle\;,\\
\label{form-s1}				
&&f^2\Sigma\,e^{-2\bar\Lambda/{T}}=
\int_0^{\omega_c}\int_0^{\omega_c} \rho (s, s^\prime )\;e^{-{s+s^\prime \over 2T}}dsds^\prime
+{1\over 24\pi}\langle\alpha_sGG\rangle \;T+{1\over 48}\;m_0^2\;\langle\bar qq\rangle\;
\end{eqnarray}
with $\rho (s,s^\prime )={1\over 24}{\alpha_s(2T)\over \pi^3} C_{mag}
\{ {s^\prime}^2 (3s-s^\prime ) \theta (s-s^\prime ) 
+ (s\leftrightarrow s^\prime ) \}$
when the currents (\ref{curr1}) and (\ref{curr2}) with no extra derivative are 
used. For $j_l^P=\displaystyle{{3\over 2}^+}$ doublet, a straightforward calculation yields
\begin{eqnarray}
\label{form-k2}
&&f^2K\,e^{-2\bar\Lambda/{T}}=-{1\over 2^7\pi^2}\int_0^{\omega_c} \omega^6\;e^{-\omega/T}d\omega+{7\over 3\times 2^5\pi}\langle\alpha_sGG\rangle \;T^3\;,\\
\label{form-s2}
&&f^2\Sigma\,e^{-2\bar\Lambda/{T}}=
\int_0^{\omega_c}\int_0^{\omega_c} \rho (s, s^\prime )\;e^{-{s+s^\prime \over 2T}}dsds^\prime
+{1\over 72\pi}\langle\alpha_sGG\rangle \;T^3\;.
\end{eqnarray}
with $\rho (s,s^\prime )={1\over 288}{\alpha_s(2T)\over \pi^3} C_{mag}
 \{ {s^\prime}^4 (s- {3\over 5}s^\prime ) \theta (s-s^\prime ) 
+ (s\leftrightarrow s^\prime ) \}$.
Combining (\ref{form-k1d})-(\ref{form-s2}) 
with (\ref{form1})-(\ref{form3}) we can
 obtain sum rules for $K$ and $\Sigma$ in the three cases.

The spin-symmetry violating term ${\cal S}$ not only causes splitting of masses in the same doublet, but also causes mixing of states with the same $j$, $P$ but different $j_l$. This mixing is characterized by the matrix element
\begin{eqnarray}
\label{mix1}
\langle j,P,j+\frac{1}{2}|\;\bar h_v\,{g\over 2}\sigma_{\mu\nu} G^{\mu\nu} h_v\;|j,P,j-\frac{1}{2}\rangle=-2\;m_{+-}(J^P)\;.
\end{eqnarray}
This quantity can be extracted from the correlator
\begin{eqnarray}
\label{mix2}
i^2\int d^4xd^4ye^{ik\cdot x-ik'\cdot y}\langle 0|T\left(J_{j,P,j+\frac{1}{2}}(x)\;\bar h_v\,{g\over 2}\sigma_{\mu\nu} G^{\mu\nu} h_v(0)\;J_{j,P,j-\frac{1}{2}}^{\dag}(y)\right )|0\rangle\;,
\end{eqnarray}
the double pole term of which is
\begin{eqnarray}
\label{mixpole}
-{2m_{+-}(J^P)\;f_{P,j+\frac{1}{2}}f_{P,j-\frac{1}{2}}\over (2\bar\Lambda_{+,j+\frac{1}{2}}-\omega)(2\bar\Lambda_{+,j-\frac{1}{2}}-\omega')}\;.
\end{eqnarray}
With the methods similar to that used above one can 
obtain the sum rule of $m_{+-}(J^P)$.
For two $1^+$ states we find
\begin{eqnarray}
\label{cross-m}
m_{\frac{1}{2},\frac{3}{2}}(1^+)\;f'_{+\frac{1}{2}}f_{+\frac{3}{2}}
e^{-\bar\Lambda_{+\frac{1}{2}}/{T}}e^{-\bar\Lambda_{+\frac{3}{2}}/{T}}=
\int_0^{\omega_c}\int_0^{\omega_c} \rho (s, s^\prime )\;e^{-{s+s^\prime \over 2T}}dsds^\prime
+{\sqrt{6}\over 72\pi}\;\langle\alpha_sGG\rangle \;T^3\;.
\end{eqnarray}
with $\rho (s,s^\prime )={\sqrt{6}\over 576 }{\alpha_s(2T)\over \pi^3} C_{mag}
\{ s{s^\prime}^4 \theta (s-s^\prime ) 
+ s^3 (6{s^\prime}^2-8ss^\prime+3s^2) \theta (s^\prime -s)  \}$
for the currents (\ref{curr4}) and (\ref{curr5}). 

The sum rules (32)-(37) and (41) were obtained in \cite{huang}
except that all loop corrections of the order $\alpha_s$ were neglected
there. In this work we have included the perturbative terms of the order
$\alpha_s$ in the sum rules for $\Sigma$ and $m_{1/2,3/2}$ which are leading
perturbative terms for these quantities.

The masses of two $1^+$ states are obtained by diagonalizing the mass matrix
\begin{eqnarray}
\label{matrix}
\left(
\begin{array}{cc}
\bar\Lambda_{+,\frac{1}{2}}-\displaystyle{{1\over 4m_Q}}
(K_{+,\frac{1}{2}}-C_{mag}\Sigma_{+,\frac{1}{2}}) & \displaystyle{\frac{1}{4m_Q}}\;C_{mag}m_{\frac{1}{2},\frac{3}{2}}(1^+)\\[4mm]
\displaystyle{\frac{1}{4m_Q}}\;C_{mag}m_{\frac{3}{2},\frac{1}{2}}(1^+) & \bar\Lambda_{+,\frac{3}{2}}-\displaystyle{{1\over 4m_Q}}(K_{+,\frac{3}{2}}+5\;C_{mag}\Sigma_{+,\frac{3}{2}})
\end{array}\;\right)\;.
\end{eqnarray}
where $C_{mag}=(\frac{\alpha_s(m_Q)}{\alpha_s(2T)})^{3\over\beta_0}$,
$\beta_0 =11-{2\over 3}n_f$. In our analysis we 
keep four active flavors with the QCD parameter ${\bar \Lambda}_{\mbox{QCD}}
=220$MeV. 

We can minimize the dependence of the sum rules obtained in 
Section \ref{sec3} on $f$, $\bar\Lambda$ and $\omega_c$
through dividing the sum rules in (\ref{form-k1d})-(\ref{form-s2}) 
by the sum rules in (\ref{form1})-(\ref{form3}) respectively.
The seemingly stability at large $T$ values is not useful, since in this 
region the sum rules is strongly contaminated by higher resonance 
states. This continuum model contamination problem is quite severe also
in the case of sum rules analysis of ground state heavy meson to $1/m_Q$
\cite{ball,neubert}. It originates from the high power dependence of the 
spectral densities. Moreover the stability plateau for the three-point 
sum rules does not necessarily coincide with that of two-point sum rules.
We follow \cite{ball} and use the same 
working windows as those of two-point sum rules
in the evaluation of ${\cal O}(1/m_Q)$ corrections.

Taking the working regions of the sum rules as that for the 
two-piont functions for the three cases, we obtain the set of values for ${\cal K}$ as following:
\begin{eqnarray}
\label{result-k1}
K=-2.30\pm 0.40 ~~\mbox{GeV}^2\;,
\end{eqnarray} 
for the doublet $(0^+,1^+)$ when the currents (\ref{curr1}), (\ref{curr2}) without the derivative are used,
\begin{eqnarray}
\label{result-k2}
K=-1.70\pm 0.40 ~~\mbox{GeV}^2,
\end{eqnarray}
for the same doublet when the currents (\ref{curr3}), (\ref{curr4}) with
the derivative are used, and
\begin{eqnarray}
\label{result-k3}
K=-1.60\pm 0.40 ~~\mbox{GeV}^2,
\end{eqnarray}
for the doublet $(1^+,2^+)$.

The results for $\Sigma$ for the doublet
$(0^+,1^+)$ in the working regions of the sum rules are
\begin{eqnarray}
\label{result-s1}
\Sigma=0.22\pm 0.04 ~~\mbox{GeV}^2\;,
\end{eqnarray} 
when the currents (\ref{curr1}) (\ref{curr2}) without the derivative 
are used and
\begin{eqnarray}
\label{result-s2}
\Sigma=0.18\pm 0.03 ~~\mbox{GeV}^2\;,
\end{eqnarray}
when the currents (\ref{curr3}), (\ref{curr4}) with
the derivative are used. For the doublet $(1^+,2^+)$, the value of $\Sigma$
is found to be
\begin{eqnarray}
\label{result-s3}
\Sigma=0.03\pm 0.01 ~~\mbox{GeV}^2\;.
\end{eqnarray}

Dividing the sum rule in (\ref{cross-m}) by the sum rules 
in (\ref{form2}) and
(\ref{form3}), we also obtain the 
expression for $m_{\frac{1}{2},\frac{3}{2}}(1^+)$. 
Using the working windows as $\omega_c=2.8-3.2$ GeV 
and $T=0.53-0.73$ GeV, we obtain
\begin{eqnarray}
\label{result-m}
m_{\frac{1}{2},\frac{3}{2}}(1^+)=0.08\pm 0.02 ~~\mbox{GeV}^2\;.
\end{eqnarray}
The errors quoted above reflect only the variations with the Borel parameter $T$ and the continuum
threshold $\omega_c$ within the working windows. They do not include other
intrinsic errors in QCD sum rules in HQET.
The corrections to the masses due to the mixing are formally of the order
$\frac{1}{m_Q^2}$. By diagonalizing the mass matrix we found that they
are numerically negligible.
								
Since the experimental data is not yet available for the excited $B$ mesons, 
we present our results for the $D$ system assuming the HQET is good enough
for the excited $D$ meson too. For the doublet $(0^+,1^+)$, we have
\begin{eqnarray}
\label{final-1}
{1\over 4}\;(m_{D_0^*}+3m_{D'_1})&=&m_c+\bar\Lambda+{1\over m_c}\;[(0.57\pm 0.10)~\mbox{GeV}^2]\;,\\
m_{D'_1}-m_{D_0^*}&=&{1\over m_c}\;[(0.22\pm 0.04)~\mbox{GeV}^2]\;,
\end{eqnarray}
when the currents (\ref{curr1}) and (\ref{curr2}) are used and
\begin{eqnarray}
\label{final-1d}
{1\over 4}\;(m_{D_0^*}+3m_{D'_1})&=&m_c+\bar\Lambda+{1\over m_c}\;[(0.42\pm 0.10)~\mbox{GeV}^2]\;,\\
m_{D'_1}-m_{D_0^*}&=&{1\over m_c}\;[(0.18\pm 0.03)~\mbox{GeV}^2]\;,
\end{eqnarray}
 when the currents (\ref{curr3}) and (\ref{curr4}) are used. As for the doublet
$(1^+,2^+)$, the result is
\begin{eqnarray}
\label{final-2}
{1\over 8}\;(3m_{D_1}+5m_{D_2^*})&=&m_c+\bar\Lambda+{1\over m_c}\;[(0.40\pm 0.10)~\mbox{GeV}^2]\;,\\
m_{D_2^*}-m_{D_1}&=&{1\over m_c}\;[(0.06\pm 0.02)~\mbox{GeV}^2]\;.
\end{eqnarray}
The renormalization coefficient $C_{mag}$ of the chromomagnetic 
operator for charmed meson in the above formulas is reasonably neglected.
The results for $B$ system are obtained by replacing $m_c$ by $m_b$ and 
multiplying the right hand sides of (51), (53) and (55) by $0.8$ 
since $C_{mag} \approx 0.8$ for B system.

The results in (\ref{final-2}) with the $\bar\Lambda$ value in
(\ref{result3}) and $m_c =1.4$GeV agree reasonably with experimental
data, though the uncertainties, especially for $\bar\Lambda$, are 
somewhat large.

%%%%%%%%%%%%%%%%%%%%%%%%%%%%%%%%%%%%%%%%%%%%%%%%%%%%%%%%%%%%%%%%%%%%%%%%%%%%%%%%%%%%%
%%%\section{The ${\cal O}(1/m_Q)$ correction to the coupling 
%%%constant $f_{1,+,{3\over 2}}$ and $\delta f_{{1\over 2},{3\over 2}}$}
%%%\label{sec8}
%%%%%%%%%%%%%%%%%%%%%%%%%%%%%%%%%%%%%%%%%%%%%%%%%%%%%%%%%%%%%%%%%%%%%%%%%%%%%%%%%%%%%
In the following sections we will need the values of 
$r_{1,+,{3\over 2}}={\delta f_{1,+, {3\over 2}}/f_{1,+, {3\over 2}} }$,  
$r_{1,-,{1\over 2}}={\delta f_{1,-, {1\over 2}}/f_{1,-, {1\over 2}} }$ and 
$\delta f_{{1\over 2},{3\over 2}}$, 
where $\delta f$ is the correction to $f$ due to the ${\cal K}$ and ${\cal S}$ 
terms in the Lagrangian at the order of ${\cal O}(1/m_Q)$ and 
$\delta f_{{1\over 2},{3\over 2}}$ is defined as:
\begin{equation}\label{deltaf}
\langle 0|J_{1,+,{1\over 2}}|1,+,{3\over 2}\rangle 
=\delta f_{{1\over 2},{3\over 2}}\eta_{1,+,{3\over 2}}\, 
\end{equation}
which is also of the order $1/m_Q$.

Omitting the indices $j, P, j_l$, $r$ can be written in the form:
\begin{equation}\label{delta-a}
r = {G_K\over 2m_Q} +d_M {G_\Sigma \over 2 m_Q}
\; ,
\end{equation}
where $G_K$ and $G_\Sigma$ are defined as:
\begin{equation}\label{gk}
\langle 0| i\int d^4x {\cal K}(x) J_{j, p, j_l}(0) |j, p, j_l\rangle =
f_{j, p, j_l}G_K\eta_{j, p, j_l} \;, 
\end{equation}
\begin{equation}\label{gsigma}
\langle 0| i\int d^4x {\cal S}(x) J_{j, p, j_l}(0) |j, p, j_l\rangle =
f_{j, p, j_l}d_MG_\Sigma\eta_{j, p, j_l} \;. 
\end{equation}

The mixing coupling constant $\delta f_{{1\over 2},{3\over 2}}
=G_{{1\over 2},{3\over 2}}/2m_Q$ where 
$G_{{1\over 2},{3\over 2}}$ is defined as: 
\begin{equation}\label{gmix}
\langle 0| i\int d^4x {\cal S}(x) J_{1,+,{1\over 2}}(0) |1,+,{3\over 2} \rangle =
G_{{1\over 2},{3\over 2}} \eta_{1,+,{3\over 2}} \; .
\end{equation}

$r_{1,-,{1\over 2}}$ for the ground state $1^-$ has been calculated in 
\cite{neub92,ball-deltaf}. Notice that $\delta f_{1,-,{1\over 2}}$ calculated in 
\cite{neub92,ball-deltaf} contains an additional term proportional to ${\bar {\bar \Lambda}}/m_Q$.
This term comes from the ${\cal O}(1/m_Q)$ correction in the expansion of the physical 
vector current ${\bar Q}\gamma_\mu q$ responsible to the leptonic decay of $B^\ast$, 
where $Q$ is the field in full QCD. For our purpose instead the matrix elements of the 
interpolating currents $J_{j,p,j_l}$ defined in (\ref{decay}) and (\ref{corr}) at 
the leading order are needed. Therefore this term is not included in (\ref{delta-a}). 
>From the results of \cite{ball-deltaf}, we know 
$r_{1,-,{1\over 2}}=-(0.82\pm 0.30)/ m_Q$ for $1^-$ 
and $r_{0,-,{1\over 2}}=-(0.74\pm 0.30)/ m_Q$ for $0^-$
heavy meson at the leading order of $\alpha_s$ \cite{ball-deltaf}. 

We follow the method used in \cite{ball-deltaf} to derive the $G_K$, $G_\Sigma$
and $G_{{1\over 2},{3\over 2}}$ for 
the $B_1$ meson and $B_2^\ast$ doublet.
Letting $k=  k^\prime$ and $\omega =\omega^\prime $ in (\ref{delta})-(\ref{poles}) 
and (\ref{mix2})-(\ref{mixpole}) we arrive at the following sum rules:
\begin{equation}\label{gk-b}
f_{+, {3\over 2}}^2 e^{-{2\bar\Lambda_{+,{3\over 2}}\over T}} 
(2G_K +{K\over T}) =-{3\over 64\pi^2}\int_0^{\omega_c} s^5 e^{-{s\over T}} ds
+{7\over 96} \langle {\alpha_s\over \pi} G^2\rangle T^2\; ,
\end{equation}
\begin{equation}\label{gs-b}
f_{+, {3\over 2}}^2 e^{-{2\bar\Lambda_{+,{3\over 2}}\over T}} 
(2G_\Sigma +{\Sigma\over T}) ={\alpha_s\over 32\pi^3}
\int_0^{\omega_c} s^5 ({25\over 81} -{4\over 15} \ln {s\over \mu}) 
e^{-{s\over T}} ds
+{1\over 72} \langle {\alpha_s\over \pi} G^2\rangle T^2\; ,
\end{equation}
\begin{equation}\label{gm-b}
f_{+, {3\over 2}}f'_{+, {1\over 2}} 
e^{-{\bar\Lambda_{+,{3\over 2}}+\bar\Lambda_{+,{1\over 2}}\over T}} 
({2G_{{1\over 2},{3\over 2}}\over f_{+, {1\over 2}} }
+{m_{+-}\over T}) ={\sqrt{6}\alpha_s\over 32\pi^3}
\int_0^{\omega_c}  s^5 ({4\over 81} -{1\over 9} \ln {s\over \mu}) 
e^{-{s\over T}} ds
+{\sqrt{6}\over 72} \langle {\alpha_s\over \pi} G^2\rangle T^2 \; ,
\end{equation}
where we take $\mu =2T$ as in the evaluation of $K$ and $\Sigma$.

In order to obtain $G_K$, $G_\Sigma$ and $G_{{1\over 2},{3\over 2}}$ 
we subtract (\ref{gk-b}), (\ref{gs-b}) and (\ref{gm-b}) by the 
expressions of (36), (37) and (41) after they are divided by $T$.
The dependence of $G_K$, $G_\Sigma$ and 
$G_{{1\over 2},{3\over 2}}/f_{+, {1\over 2}}$ 
on the parameter $T$ and $\omega_c$ is 
shown in Fig. 1-3 with $\omega_c=3.2, 3.0, 2.8$ GeV. 
In the numerical analysis we choose the same 
stability region of $(T, \omega_c)$ as those from the mass sum rules.
In this region there exists stability plateu for the sum rules 
(\ref{gk-b})-(\ref{gs-b}). For the sum rule (\ref{gm-b}) the gluon 
condensate contribution is much bigger than the perturbative one, which 
results in the curves in Fig. 3. 

Numerically we have: 
$G_K=-(1.0\pm 0.20\pm 0.25)$GeV, $G_\Sigma =(0.013\pm 0.005\pm 0.002)$GeV,
Here the first error comes from the variation with $\omega_c$ and the 
second error is due to the variation with $T$ in the working region.
$G_{{1\over 2},{3\over 2}}/ f_{+, {1\over 2}} =(0.023\pm 0.003\pm 0.010)$GeV.
Finally we arrive at $r_{1,+,{3\over 2}}=-(0.52\pm 0.22)/m_Q$, 
$r_{2,+,{3\over 2}}=-(0.57\pm 0.25)/ m_Q$, 
$\delta f_{{1\over 2},{3\over 2}}=(4.3\pm 2.0)\times 10^{-3}/m_Q$GeV$^{5/2}$.

%%%%%%%%%%%%%%%%%%%%%%%%%%%%%%%%%%%%%%%%%%%%%%%%%%%%%%%%%%%%%%%%%%%%%%%%%%%%%%%%%%%
%%%%%%%%%%%%%%%%%%%%%%%%%%%%%%%%%%%%%%%%%%%%%%%%%%%%%%%%%%%%%%%%%%%%%%%%%%%%%%%%%%%
\section{Mixing of the Two $1^+$ States}
\label{sec4}
Diagonalizing the mixing mass matrix (\ref{matrix}) 
we arrive at the physical eigenstates for the two $1^+$ states up to
the first order of $1/m_Q$. 
\begin{equation}\label{mix-state1}
|B'_1\rangle =\cos\theta |1,+,{1\over 2}\rangle
-\sin\theta |1,+,{3\over 2}\rangle \, ,
\end{equation}
\begin{equation}\label{mix-state2}
|B_1 \rangle =\sin\theta |1,+,{1\over 2}\rangle
+\cos\theta |1,+,{3\over 2}\rangle \, ,
\end{equation}
where $\theta$ is the mixing angle, which is determined by the 
folowing equation:
\begin{equation}\label{theta}
\tan 2\theta =-\frac{2C_{mag} m_{{1\over 2},{3\over 2}} }{4m_Q(
{\bar  \Lambda}_{+,{1\over 2}}-{\bar  \Lambda}_{+,{3\over 2}})
+(K_{+,{1\over 2}}-K_{+,{3\over 2}}) 
+C_{mag}(\Sigma_{+,{1\over 2}}+5\Sigma_{+,{3\over 2}})}\;.
\end{equation}
Numerically we get $\theta =-(0.10\pm 0.05)$ for the $D$ system with $m_c=1.4$GeV and 
$\theta =-(0.03\pm 0.015)$ for the $B$ system with $m_b=4.8$GeV. 

The corresponding interpolating 
currents which annihiliate one of these two states and do not couple to
the other up to the first order of $1/m_Q$ can be written as 
\begin{equation}\label{mix-current1}
J^{B'_1}=J_{1,+,{1\over 2}} +\alpha J_{1,+,{3\over 2}} \, ,
\end{equation}
\begin{equation}\label{mix-current2}
J^{B_1}=J_{1,+,{3\over 2}} +\beta J_{1,+,{1\over 2}} \, .
\end{equation}
In the discussion of the mixing of the two $1^+$ states we always 
use the current (\ref{curr4}) with derivatives.  
The mixing parameters $\alpha$, $\beta$ are determined from the orthonormal 
conditions of the physical $1^+$ states:
\begin{equation}\label{mix-ortho1}
\langle 0| J^{B'_1} | B_1\rangle =0 \, ,
\end{equation}
\begin{equation}\label{mix-ortho2}
\langle 0| J^{B_1} | B'_1 \rangle =0 \, .
\end{equation}
Solving the above equations we obtain
\begin{equation}\label{alpha}
\alpha = -{ f_{1,+,{1\over 2}} \tan\theta  +
\delta f_{{1\over 2},{3\over 2}}\over f_{1,+,{3\over 2}}} \,
\end{equation}
\begin{equation}\label{beta}
\beta = { f_{1,+,{3\over 2}} \tan\theta
- \delta f_{{3\over 2},{1\over 2}}\over f_{1,+,{1\over 2}}} \,
\end{equation}

With the values of f's in (\ref{ff1})-(\ref{ff3}), 
$\delta f_{{1\over 2},{3\over 2}}$ and $\tan\theta$ we get 
$\alpha =0.22\pm 0.10$, $\beta =-(0.06\pm 0.03)$ for $D'_1$, $D_1$ and 
$\alpha =0.065\pm 0.032$, $\beta =-(0.02\pm 0.01)$ for $B'_1$, 
$B_1$ respectively.
%%%%%%%%%%%%%%%%%%%%%%%%%%%%%%%%%%%%%%%%%%%%%%%%%%%%%%%%%%%%%%%%%%%%%%%%%%%%%%%%%%
%%%%%%%%%%%%%%%%%%%%%%%%%%%%%%%%%%%%%%%%%%%%%%%%%%%%%%%%%%%%%%%%%%%%%%%%%%%%%%%%%%
\section{Pionic decay amplitudes of $B_1$ 
up to the order of ${\cal O}(1/m_Q)$ }
\label{sec5}
In this section we shall use the light-cone QCD sum rules (LCQSR) \cite{bely95}
to calculate the partial widths of the pionic decay processes of 
$B_1$ and $B_2^\ast$. The decay width of $(1^+,2^+)$, $(0^+,1^+)$ doublets 
have been calculated at the leading order in HQET with conventional 
QCD sum rules \cite{ybd97}. Different from the conventional QCD sum rules, 
which is based on the short distance operator product expansion (OPE), 
LCQSR is based on the OPE on the light cone which is the  
expansion over the twists of the operators. 

Denote the doublet $(1^+,2^+)$ by $(B_1,B_2^*)$.
Let us consider first the decay process $B_1\to B^\ast\pi$. 
>From (\ref{mix-state2}) the decay amplitude can be written as 
\begin{equation}\label{amp-1}
M(B_1\to B^\ast\pi )=\cos\theta M({3\over 2}^+)+\sin\theta M({1\over 2}^+) \; ,
\end{equation}
where $M({3\over 2}^+)$ and $M({1\over 2}^+)$ denote contributions from the 
states $j_{\ell}={3\over 2}^+$ and $j_{\ell}={1\over 2}^+$ respectively.
Up to the order of ${\cal O}(1/m_Q)$ we can confine us to the leading order 
for $M({1\over 2}^+)$ and up to the next-to-leading order for 
for $M({3\over 2}^+)$ in the $1/m_Q$ expansion.

>From covariance and conservation of the angular momentum of 
the light component in the $m_Q\to\infty$ limit, the leading 
order amplitudes have the general form: 
\begin{equation}
\label{coup1}
 M_0(B_1\to B^*\pi)=M_0({3\over 2}^+)=
 I\;\epsilon^*_{\mu}\eta_{\nu}(q_t^{\mu}q_t^{\nu}-\frac{1}{3}g_t^{\mu\nu}
 q_t^2)g(B_1,B^*)\;,
\end{equation} 
\begin{equation}
\label{coup2}
M({1\over 2}^+)=I\;\epsilon^*\cdot\eta 
 g^\prime (B_1', B^* )\; ,
\end{equation}
corresponding to the D-wave and S-wave for the pion respectively.
Here $\eta_{\mu}$ and $\epsilon_{\mu}$ are polarization
vectors for the states $1^+$ and $1^-$ respectively.
$q_{t\mu}=q_{\mu}-v\cdot qv_{\mu}$. $I=\sqrt{2}$, $1$ 
for charged and neutral pion respectively.

The ${\cal O}(1/m_Q)$ correction $M_1 ({3\over 2}^+)$ arises from the 
the terms ${\cal K}$ and ${\cal S}$ in the Lagrangian in HQET. The operator ${\cal K}$ 
contributes only to the D-wave amplitude, while ${\cal S}$ contributes to both the 
D-wave and S-wave amplitudes. Therefore the total ${\cal O}(1/m_Q)$ correction 
of the decay amplitude $M(B_1\to B^*\pi)$ has the general form:
\begin{eqnarray}
\label{coup3}
 M_1(B_1\to B^*\pi)&=M_{1d}+M_{1s}=&I\;\epsilon^*_{\mu}\eta_{\nu}
 \{ (q_t^{\mu}q_t^{\nu}-\frac{1}{3}g_t^{\mu\nu}q_t^2 )g_{1d}
 +g_t^{\mu\nu}  g_{1s}\} \;.
\end{eqnarray} 

In order to get a consistent expansion over $1/m_Q$ we consider the correlator 
\begin{eqnarray}
\label{6a}
 \int d^4x\;e^{-ik\cdot x}\langle\pi(q)|T\left(J^{\beta}_{1,-,\frac{1}{2}}(0)
 J^{\dagger\alpha}_{B_1}(x)\right)|0\rangle \; ,
\end{eqnarray}
where $J_{B_1}$ is the interpolating current (\ref{mix-current2}) for $B_1$. 
It is implied that the ${\cal O}(1/m_Q)$ terms in the Lagrangian of HQET
are included in the calculation of (\ref{6a}).

The pole term of (\ref{6a}) can be expressed up to the order ${\cal O}(1/m_Q)$ as:
\begin{eqnarray}
\label{6c}\nonumber
{(f_{-,{1\over 2}} +\delta f_{-,{1\over 2}})
(f_{+,{3\over 2}}+ \delta f_{+,{3\over 2}} )
(M_0 + M_{1d} +M_{1s} )^{\alpha\beta}
\over 
[2(\bar\Lambda_{-,{1\over 2}}+\delta m_{-,{1\over 2}}) -\omega']
[2(\bar\Lambda_{+,{3\over 2}}+\delta m_{+,{3\over 2}}) -\omega]
}  &\\ 
+{d\over 2(\bar\Lambda_{+,{3\over 2}}+\delta m_{+,{3\over 2}}) -\omega}
+{d'\over 2(\bar\Lambda_{-,{1\over 2}}+\delta m_{-,{1\over 2}}) -\omega'] }
\;,
\end{eqnarray}
which can be expanded as:
\begin{eqnarray}
\label{6d}\nonumber
{f_{-,{1\over 2}}f_{+,{3\over 2}}M_0^{\alpha\beta}
\over (2\bar\Lambda_{-,{1\over 2}} -\omega')(2\bar\Lambda_{+,{3\over 2}}-\omega)} 
+{f_{-,{1\over 2}}f_{+,{3\over 2}}(r_{1,-,{1\over 2}} +r_{1,+,{3\over 2}})M_0^{\alpha\beta}
\over (2\bar\Lambda_{-,{1\over 2}} -\omega')(2\bar\Lambda_{+,{3\over 2}}-\omega)} 
+{f_{-,{1\over 2}}f_{+,{3\over 2}}(M_{1d} +M_{1s})^{\alpha\beta}
\over (2\bar\Lambda_{-,{1\over 2}} -\omega')(2\bar\Lambda_{+,{3\over 2}}-\omega)} 
&\\  
-{f_{-,{1\over 2}}f_{+,{3\over 2}}2\delta m_{1\over 2} M_0^{\alpha\beta}
\over (2\bar\Lambda_{-,{1\over 2}} -\omega')^2 (2\bar\Lambda_{+,{3\over 2}}-\omega)} 
-{f_{-,{1\over 2}}f_{+,{3\over 2}}2\delta m_{3\over 2}M_0^{\alpha\beta}
\over (2\bar\Lambda_{-,{1\over 2}} -\omega')(2\bar\Lambda_{+,{3\over 2}}-\omega)^2} 
+\cdots 
\;,
\end{eqnarray}
where $M_i^{\alpha\beta}$, $i=0, 1$, is defined by $M_i=\epsilon^\ast_\alpha
\eta_\beta M_i^{\alpha\beta}$, and $r_{1,-,{1\over 2}}$, $r_{1,+,{3\over 2}}$ 
is defined in Section (\ref{sec4}).

On the other hand (\ref{6a}) can be expanded as:
\begin{eqnarray}
\label{6b}\nonumber
 \int d^4x\;e^{-ik\cdot x}\langle\pi(q)|T\left(J^{\beta}_{1,-,\frac{1}{2}}(0)
 J^{\dagger\alpha}_{1,+,\frac{3}{2}}(x)\right)|0\rangle  &\\ \nonumber
+\beta \int d^4x\;e^{-ik\cdot x}\langle\pi(q)|T\left(J^{\beta}_{1,-,\frac{1}{2}}(0)
 J^{\dagger\alpha}_{1,+,\frac{1}{2}}(x)\right)|0\rangle  &\\ \nonumber
+ i\int d^4xd^4y\;e^{-ik\cdot x}e^{ik^\prime \cdot y} 
\langle\pi(q)|T\left(J^{\beta}_{1,-,\frac{1}{2}}(y)
 {g_s \over 4m_Q}{\bar h}_v(0) \sigma\cdot G h_v(0) 
 J^{\dagger\alpha}_{1,+,\frac{3}{2}}(x)\right)|0\rangle  &\\ 
+ i\int d^4xd^4y\;e^{-ik\cdot x}e^{ik^\prime \cdot y} 
\langle\pi(q)|T\left(J^{\beta}_{1,-,\frac{1}{2}}(y)
 {1 \over 2m_Q}{\bar h}_v(0) (iD_t)^2 h_v(0) 
 J^{\dagger\alpha}_{1,+,\frac{3}{2}}(x)\right)|0\rangle 
\; ,
\end{eqnarray}
where the first term is the correlator at the leading order of $1/m_Q$ expansion, 
the second term is due to the mixing of two $1^+$ currents, the third and fourth term 
arises from the ${\cal O}(1/m_Q)$ corrections to the transition 
$j_{\ell}={3\over 2}^+$ to $j_{\ell}={1\over 2}^-$ due to the 
${\cal O}(1/m_Q)$ terms of the Lagrangian in the order of ${\cal O}(1/m_Q)$. 
These correlators will be calculated with OPE and 
the leading order Lagrangian of HQET.

%%%%%%%%%%%%%%%%%%%%%%%%%%%%%%%%%%%%%%%%%%%%%%%%%%%%%%%%%%%%%%%%%%%%%%%%%%%%%%%%%%%%%%%%
%%%%%%%%%%%%%%%%%%%%%%%%%%%%%%%%%%%%%%%%%%%%%%%%%%%%%%%%%%%%%%%%%%%%%%%%%%%%%%%%%%%%%%%%
\section{The sum rules for the coupling constants $g$, $g'$ etc }
\label{sec6}

We consider the correlators appearing in (\ref{6b}). They have the general form:
\begin{eqnarray}
\label{7a}
 \int d^4x\;e^{-ik\cdot x}\langle\pi(q)|T\left(J^{\beta}_{1,-,\frac{1}{2}}(0)
 J^{\dagger\alpha}_{1,+,\frac{3}{2}}(x)\right)|0\rangle = 
 \left(q_t^{\alpha}q_t^{\beta}-\frac{1}{3}g_t^{\alpha\beta}q^2_t\right)G(\omega,\omega')\;,
\end{eqnarray}
\begin{eqnarray}
\label{7d}
\int d^4x\;e^{-ik\cdot x}\langle\pi(q)|T\left(J^{\beta}_{1,-,\frac{1}{2}}(0)
 J^{\dagger\alpha}_{1,+,\frac{1}{2}}(x)\right)|0\rangle = 
 g_t^{\alpha\beta} G' (\omega,\omega')\;,
\end{eqnarray}
\begin{eqnarray}
\label{7b}
i \int d^4xd^4y\;e^{-ik\cdot x}e^{ik^\prime \cdot y} 
 \langle\pi(q)|T\left(J^{\beta}_{1,-,\frac{1}{2}}(y),
 {g_s\over 4} {\bar h}_v (0) \sigma\cdot G h_v (0), 
 J^{\dagger\alpha}_{1,+,\frac{3}{2}}(x)\right)|0\rangle \\ \nonumber
 = m_Q \left(\left(q_t^{\alpha}q_t^{\beta}-\frac{1}{3}
 g_t^{\alpha\beta}q^2_t\right)G^S_{1d} (\omega,\omega')
+g_t^{\alpha\beta} G^S_{1s} (\omega,\omega')\right)
\;,
\end{eqnarray}
\begin{eqnarray}
\label{7c}\nonumber
 i\int d^4xd^4y\;e^{-ik\cdot x}e^{ik^\prime \cdot y} 
 \langle\pi(q)|T\left(J^{\beta}_{1,-,\frac{1}{2}}(y),
 {1\over 2} {\bar h}_v (0) (iD_t)^2 h_v (0), 
 J^{\dagger\alpha}_{1,+,\frac{3}{2}}(x)\right)|0\rangle \\ 
 = m_Q \left(q_t^{\alpha}q_t^{\beta}-\frac{1}{3}
 g_t^{\alpha\beta}q^2_t\right)G^K_{1d} (\omega,\omega')
\;,
\end{eqnarray}
where $k^{\prime}=k-q$, $\omega=2v\cdot k$, $\omega^{\prime}=2v\cdot
k^{\prime}$ and $q^2=m_\pi^2 \approx 0$. 
The suffices ${\cal K}$ and ${\cal S}$ denote the contributions from the operators 
from ${\cal K}$ and ${\cal S}$ respectively.
The forms of the right hand side of (\ref{7a})-(\ref{7d})
are determined by that $\alpha$ and $\beta$ are transverse
indices, $x-y=vt$ on the heavy quark propagator and the 
conservation of angular momentum of the light component.

For deriving QCD sum rules we first calculate the correlator (\ref{7a}),
which is equal to 
\begin{equation}\label{lam5}			
{\sqrt{6}\over 8} \int_0^{\infty} dt \int dx e^{-ikx} 
\delta (-x-vt){\bf Tr} \{ 
 \gamma^t_\beta (1+{\hat v})\gamma_5 (D^t_\alpha 
 -{1\over 3}\gamma^t_\alpha {\hat D}^t )
\langle \pi (q)|u(0) {\bar d}(x) |0\rangle  
 \} \; ,
\end{equation}

By the operator expansion on the light-cone
the matrix element of the nonlocal operators between the vacuum and 
pion state in (\ref{lam5}) defines the two particle wave function of the pion. 
Up to twist four the Dirac components of this wave  function can be 
written as \cite{bely95}:
\begin{eqnarray}
<\pi(q)| {\bar d} (x) \gamma_{\mu} \gamma_5 u(0) |0>&=&-i f_{\pi} q_{\mu} 
\int_0^1 du \; e^{iuqx} (\varphi_{\pi}(u) +x^2 g_1(u) + {\cal O}(x^4) ) 
\nonumber \\
&+& f_\pi \big( x_\mu - {x^2 q_\mu \over q x} \big) 
\int_0^1 du \; e^{iuqx}  g_2(u) \hskip 3 pt  , \label{phipi} \\
<\pi(q)| {\bar d} (x) i \gamma_5 u(0) |0> &=& {f_{\pi} m_{\pi}^2 \over m_u+m_d} 
\int_0^1 du \; e^{iuqx} \varphi_P(u)  \hskip 3 pt ,
 \label{phip}  \\
<\pi(q)| {\bar d} (x) \sigma_{\mu \nu} \gamma_5 u(0) |0> &=&i(q_\mu x_\nu-q_\nu 
x_\mu)  {f_{\pi} m_{\pi}^2 \over 6 (m_u+m_d)} 
\int_0^1 du \; e^{iuqx} \varphi_\sigma(u)  \hskip 3 pt .
 \label{phisigma}
\end{eqnarray}
\noindent 
The wave function $\varphi_{\pi}$ is associated with the leading twist two 
operator, $g_1$ and $g_2$ correspond to twist four operators, and $\varphi_P$ and 
$\varphi_\sigma$ to twist three ones. Due to the choice of the
gauge  $x^\mu A_\mu(x) =0$, the path-ordered gauge factor
$P \exp\big(i g_s \int_0^1 du x^\mu A_\mu(u x) \big)$ has been omitted.
The coefficient in front of the r.h.s. of 
eqs. (\ref{phip}), (\ref{phisigma})
can be written in terms of the light quark condensate
$<{\bar u} u>$ using the PCAC relation: 
$\displaystyle \mu_{\pi}= {m_\pi^2 \over m_u+m_d}
=-{2 \over f^2_\pi} <{\bar u} u>$. 

From (\ref{lam5}), (\ref{phipi}) and (\ref{phisigma}) we arrive at
\begin{equation}\label{quark}
G(\omega ,\omega')=
-{\sqrt{6}\over 8}if_\pi\int_0^{\infty} dt \int_0^1 du e^{i (1-u) {\omega t \over 2}}
e^{i u {\omega' t \over 2}} u \{ \varphi_{\pi} (u) +
t^2g_1(u)+{it\over q\cdot v}g_2(u)+{it\over 6}\mu_{\pi}\varphi_{\sigma}(u) \}
+\cdots\; .
\end{equation}
For large euclidean values of $\omega$ and $\omega'$ 
this integral is dominated by the region of small $t$, therefore it can be 
approximated by the first a few terms.

Similarly we have:
\begin{eqnarray}\label{quark-2}\nonumber
&G'(\omega ,\omega')=
-{i\over 4}q_t^2 f_\pi\int_0^{\infty} dt \int_0^1 du e^{i (1-u) {\omega t \over 2}}
e^{i u {\omega' t \over 2}} u \{ \varphi_{\pi} (u) +
t^2g_1(u)+{it\over q\cdot v}g_2(u)+{it\over 6}\mu_{\pi}\varphi_{\sigma}(u) \} 
+\cdots\; \\
&
= +{i\over 16}(\omega -\omega')^2
 f_\pi\int_0^{\infty} dt \int_0^1 du e^{i (1-u) {\omega t \over 2}}
e^{i u {\omega' t \over 2}} u \{ \varphi_{\pi} (u) +
t^2g_1(u)+{it\over q\cdot v}g_2(u)+{it\over 6}\mu_{\pi}\varphi_{\sigma}(u) \} 
+\cdots\; ,
\end{eqnarray}
with the current (\ref{curr4}) for $j_l ={1\over 2}^+$ state. Or  
\begin{equation}\label{quark-3}
G'(\omega ,\omega')=
{i\over 4}f_\pi\int_0^{\infty} dt \int_0^1 du e^{i (1-u) {\omega t \over 2}}
e^{i u {\omega' t \over 2}}  \{\mu_\pi \varphi_P (u) -(q\cdot v)[
\varphi_\pi (u) + t^2g_1(u)]
\}+\cdots\; .
\end{equation}
with the current (\ref{curr2}).

After double Borel transformation with the variables $\omega$ and $\omega'$
the single-pole terms in (\ref{6c}) are eliminated. Using (\ref{6b}), 
(\ref{7a}), (\ref{quark}) and
subtracting the continuum contribution which is modeled by the
dispersion integral in region $\omega ,\omega' \ge \omega_c$, we arrive at:
\begin{equation}\label{final-a}
 g f_{-,{1\over 2} } f_{+, {3\over 2} } = -{\sqrt{6}\over 4}F_{\pi}
 e^{ { {\bar \Lambda}_{-,{1\over 2} } +{\bar \Lambda}_{+,{3\over 2} } \over T }}
 \{ u_0\varphi_{\pi} (u_0)T(1-e^{-{\omega_c\over T}}) -
{4\over T}u_0g_1(u_0)+{4\over T}G_2(u_0)
+{1\over 3}\mu_{\pi}u_0\varphi_{\sigma}(u_0) \}\;,
\end{equation}
with $F_{\pi}={f_\pi\over \sqrt{2}}=92$MeV, $u_0\equiv {T_1\over T_1+T_2}$, 
$T\equiv {T_1T_2\over T_1+T_2}$, where $T_1$, $T_2$ are the Borel parameters
and $G_2 (u_0)\equiv \int_0^{u_0} ug_2(u)du$.
In obtaining (\ref{final-a}) we have used the Borel transformation formula:
${\hat {\cal B}}^T_{\omega} e^{\alpha \omega}=\delta (\alpha -{1\over T})$.

Similarly, for the coupling constant $g'$ we have:
\begin{eqnarray}\label{final-aa}\nonumber
& g' f_{-,{1\over 2} } f'_{+, {1\over 2} } = {1\over 8}F_{\pi}
 e^{ { {\bar \Lambda}_{-,{1\over 2} } +{\bar \Lambda}_{+,{1\over 2} } \over T }}
 \{
\frac{d^2}{du^2}\left( 
u\varphi_{\pi} (u)T^3f_2({\omega_c\over T})
-4ug_1(u)Tf_0({\omega_c\over T}) 
+{1\over 3}\mu_{\pi}u\varphi_{\sigma}(u)T^2f_1({\omega_c\over T}) 
\right)|_{u=u_0}\\&
+4\frac{d}{du}\left(ug_2(u)\right)|_{u=u_0}Tf_0({\omega_c\over T}) 
\}\;,
\end{eqnarray}
with the current (\ref{curr4}), 
where $f_n(x)=1-e^{-x}\sum\limits_{k=0}^{n}{x^k\over k!}$ is the factor used 
to subtract the continuum. The derivative in (\ref{final-aa}) arises from 
the factor $q^2_t$ in (\ref{quark-2}). We have used integration by parts 
to absorb the factors $q\cdot v$ and $(q\cdot v)^2$. In this way we 
arrive at the simple form after double Borel transformation. 

With the current (\ref{curr2}) we have 
\begin{equation}\label{final-bb}
 g' f_{-,{1\over 2} } f_{+, {1\over 2} } =  {1\over 4}F_{\pi}
 e^{ { {\bar \Lambda}_{-,{1\over 2} } +{\bar \Lambda}^\prime_{+,{1\over 2} } \over T }}
 \{ -\varphi_\pi^{\prime} (u_0) T^2f_1({\omega_c\over T}) +
2\mu_\pi \varphi_P (u_0)T f_0({\omega_c\over T}) +4 g'_1(u_0)
\}\;,
\end{equation}
where $\varphi_\pi^{\prime} (u_0)$, $g'_1(u_0)$ are the first derivatives of
$\varphi_\pi (u)$, $g_1(u)$ at $u=u_0$.

%%%%%%%%%%%%%%%%%%%%higher-twist wave functions%%%%%%%%%%%%%%%%%%%%%%%%%%%%%%%
For the ${\cal O}(1/m_Q)$ correction to the decay amplitude for
$(j_{\ell}={3\over 2}) \to (j_{\ell}={1\over 2}) +\pi $, we 
need to consider the correlators (\ref{7b}) and (\ref{7c}).
In calculations of these correlators enter 
the pion matrix elements of quark-gluon operators,  
which are parameterized in terms of pion wave functions 
defined in \cite{bely95}:
\begin{eqnarray}
& &<\pi(q)| {\bar d} (x) \sigma_{\alpha \beta} \gamma_5 g_s 
G_{\mu \nu}(ux)u(0) |0>=
\nonumber \\ &&i f_{3 \pi}[(q_\mu q_\alpha g_{\nu \beta}-q_\nu q_\alpha g_{\mu \beta})
-(q_\mu q_\beta g_{\nu \alpha}-q_\nu q_\beta g_{\mu \alpha})]
\int {\cal D}\alpha_i \; 
\varphi_{3 \pi} (\alpha_i) e^{iqx(\alpha_1+v \alpha_3)} \;\;\; ,
\label{p3pi} 
\end{eqnarray}

\begin{eqnarray}
& &<\pi(q)| {\bar d} (x) \gamma_{\mu} \gamma_5 g_s 
G_{\alpha \beta}(vx)u(0) |0>=
\nonumber \\
&&f_{\pi} \Big[ q_{\beta} \Big( g_{\alpha \mu}-{x_{\alpha}q_{\mu} \over q \cdot 
x} \Big) -q_{\alpha} \Big( g_{\beta \mu}-{x_{\beta}q_{\mu} \over q \cdot x} 
\Big) \Big] \int {\cal{D}} \alpha_i \varphi_{\bot}(\alpha_i) 
e^{iqx(\alpha_1 +v \alpha_3)}\nonumber \\
&&+f_{\pi} {q_{\mu} \over q \cdot x } (q_{\alpha} x_{\beta}-q_{\beta} 
x_{\alpha}) \int {\cal{D}} \alpha_i \varphi_{\|} (\alpha_i) 
e^{iqx(\alpha_1 +v \alpha_3)} \hskip 3 pt  \label{gi} 
\end{eqnarray}
\noindent and
\begin{eqnarray}
& &<\pi(q)| {\bar d} (x) \gamma_{\mu}  g_s \tilde G_{\alpha \beta}(vx)u(0) |0>=
\nonumber \\
&&i f_{\pi} 
\Big[ q_{\beta} \Big( g_{\alpha \mu}-{x_{\alpha}q_{\mu} \over q \cdot 
x} \Big) -q_{\alpha} \Big( g_{\beta \mu}-{x_{\beta}q_{\mu} \over q \cdot x} 
\Big) \Big] \int {\cal{D}} \alpha_i \tilde \varphi_{\bot}(\alpha_i) 
e^{iqx(\alpha_1 +v \alpha_3)}\nonumber \\
&&+i f_{\pi} {q_{\mu} \over q \cdot x } (q_{\alpha} x_{\beta}-q_{\beta} 
x_{\alpha}) \int {\cal{D}} \alpha_i \tilde \varphi_{\|} (\alpha_i) 
e^{iqx(\alpha_1 +v \alpha_3)} \hskip 3 pt . \label{git} 
\end{eqnarray}
\noindent 
The operator $\tilde G_{\alpha \beta}$  is the dual of $G_{\alpha \beta}$:
$\tilde G_{\alpha \beta}= {1\over 2} \epsilon_{\alpha \beta \delta \rho} 
G^{\delta \rho} $; ${\cal{D}} \alpha_i$ is defined as 
${\cal{D}} \alpha_i =d \alpha_1 
d \alpha_2 d \alpha_3 \delta(1-\alpha_1 -\alpha_2 
-\alpha_3)$. 
The function $\varphi_{3 \pi}$ is of twist three, while all the wave 
functions appearing in eqs.(\ref{gi}), (\ref{git}) are of twist four.
The wave functions $\varphi (x_i,\mu)$ ($\mu$ is the renormalization point) 
describe the distribution in longitudinal momenta inside the pion, the 
parameters $x_i$ ($\sum_i x_i=1$) 
representing the fractions of the longitudinal momentum carried 
by the quark and the antiquark. 

The wave function normalizations immediately follow from the definitions
(\ref{phipi})-(\ref{git}):
$\int_0^1 du \; \varphi_\pi(u)=\int_0^1 du \; \varphi_\sigma(u)=1$,
$\int_0^1 du \; g_1(u)={\delta^2/12}$,
$\int {\cal D} \alpha_i \varphi_\bot(\alpha_i)=
\int {\cal D} \alpha_i \varphi_{\|}(\alpha_i)=0$,
$\int {\cal D} \alpha_i \tilde \varphi_\bot(\alpha_i)=-
\int {\cal D} \alpha_i \tilde \varphi_{\|}(\alpha_i)={\delta^2/3}$,
with the parameter $\delta$ defined by 
the matrix element: 
$<\pi(q)| {\bar d} g_s \tilde G_{\alpha \mu} \gamma^\alpha u |0>=
i \delta^2 f_\pi q_\mu$.

For sum rules for the $1/m_Q$ corrections we shall compare 
the double Borel transform of (\ref{6d}) with those of (\ref{7b}) and (\ref{7c}) 
for the D-wave and S-wave part seperately. It is convenient to confine $k_\mu$ 
to be longitudinal, i.e., $k^t_\mu =0$.
For the D-wave amplitude we have:
\begin{eqnarray}\label{amp-b}\nonumber
&G^S_{1d}(\omega, \omega') =
+{\sqrt{6}\over 12}{1\over m_Q}
\int dt_1dt_2 {\cal D} \alpha_i \alpha_2
[ - {f_{3\pi}\over \sqrt{2}} (q\cdot v) \varphi_{3\pi} (\alpha_i)
+F_\pi \varphi_\bot (\alpha_i) 
+{F_\pi \over 2}\tilde\varphi_{\|} (\alpha_i) ] \\ 
& e^{ikvt}e^{-iqv[\alpha_3 t_2 +\alpha_2 (t_1+t_2)]}\; , 
\end{eqnarray}
\begin{equation}\label{amp-c}
G^S_{1s}(\omega, \omega')= {q_t^2\over 3}G^S_{1d}(\omega, \omega') \;,
\end{equation}
\begin{eqnarray}\label{amp-d}\nonumber
&G^K_{1d}(\omega ,\omega')=
{\sqrt{6}\over 16}{1\over m_Q}
F_\pi\int_0^{\infty} dt_1 dt_2 \int_0^1 
du e^{i (1-u) {\omega (t_1+t_2) \over 2}}
e^{i u {\omega' (t_1+t_2) \over 2}}(k-uq)_t^2  u \{ \varphi_{\pi} (u) \\
&+(t_1+t_2)^2g_1(u)+{i(t_1+t_2)\over q\cdot v}g_2(u)+
{i(t_1+t_2)\over 6}\mu_{\pi}\varphi_{\sigma}(u) \}
+\cdots\; ,
\end{eqnarray}
where $(k-uq)_t^2=-{u^2\over 4}(\omega -\omega')^2$.

After double Borel transformation we have:
\begin{eqnarray}\label{final-b}\nonumber
&[(r_{1,-,{1\over 2}} +r_{1,+,{3\over 2}})g + g_{1d}- 
{\delta m_{1\over 2} +\delta m_{3\over 2} \over T}g ] 
 f_{-,{1\over 2} } f_{+, {3\over 2} }\\ \nonumber
&= e^{{ {\bar \Lambda}_{-,{1\over 2} } +{\bar \Lambda}_{+,{3\over 2} } \over T }} \{
-{\sqrt{6}\over 3}{F_\pi\over m_Q}
\int {\cal D} \alpha_i {\alpha_2\over \alpha_3} 
[ \varphi_\bot (\alpha_i) 
+{1\over 2}\tilde\varphi_{\|} (\alpha_i) ] \\ 
&+{\sqrt{3}\over 6}{f_{3\pi}\over m_Q}
\int {\cal D} \alpha_i {1\over \alpha_3} 
 \frac{d}{d\alpha_2}\left( \alpha_2\varphi_{3\pi} (\alpha_i)\right)
  Tf_0({\omega_c\over T})
+{\sqrt{6}\over 16}{F_{\pi}\over m_Q}K(u_0,T)
\}\;, 
\end{eqnarray}
where the function $K(u_0,T)$ arises from the kinetic operator ${\cal K}$, which is 
defined as:
\begin{eqnarray}\label{KKK}\nonumber
&K(u_0,T) \equiv 
\frac{d^2}{du^2}\left( 
u^3\varphi_{\pi} (u)T^2f_1({\omega_c\over T})
-4u^3g_1(u)
+{1\over 3}\mu_{\pi}u^3\varphi_{\sigma}(u)Tf_0({\omega_c\over T}) 
\right)|_{u=u_0}\\ &
+4\frac{d}{du}\left(u^3g_2(u)\right)|_{u=u_0}
\;.
\end{eqnarray}
In order to extract $g_{1d}$ we have to use  
the operator ${d\over dT}\{T\times (\ref{final-b}) \}$.

For the S-wave part we find 
\begin{eqnarray}\label{final-d}\nonumber
& g_{1s} f_{-,{1\over 2} } f_{+, {3\over 2} } 
 = e^{ { {\bar \Lambda}_{-,{1\over 2} } +{\bar \Lambda}_{+,{3\over 2} } \over T }} \{
-{\sqrt{6}\over 9}{F_\pi\over m_Q} q_t^2
\int {\cal D} \alpha_i {\alpha_2\over \alpha_3} 
[ \varphi_\bot (\alpha_i) 
+{1\over 2}\tilde\varphi_{\|} (\alpha_i) ] \\ \nonumber
&+{\sqrt{3}\over 18}{f_{3\pi}\over m_Q} q_t^2
\int {\cal D} \alpha_i {1\over \alpha_3} 
 \frac{d}{d\alpha_2}\left( \alpha_2\varphi_{3\pi} (\alpha_i)\right)
  Tf_0({\omega_c\over T}) \\ \nonumber
&+{\beta\over 8}F_{\pi}\{
\frac{d^2}{du^2}\left( 
u\varphi_{\pi} (u)T^3f_2({\omega_c\over T})
-4ug_1(u)Tf_0({\omega_c\over T})
+{1\over 3}\mu_{\pi}u\varphi_{\sigma}(u)T^2f_1({\omega_c\over T}) 
\right)|_{u=u_0}\\ 
&+4\frac{d}{du}\left(ug_2(u)\right)|_{u=u_0}Tf_0({\omega_c\over T})
\} \; ,
\end{eqnarray}
where $\beta$ is defined in section \ref{sec4} and its value is found 
through (\ref{beta}).
%%%%%%%%%%%%%%%%%%%%%%%%%%%%%%%%%%%%%%%%%%%%%%%%%%%%%%%%%%%%%%%%%%%%%%%%%%%%%%%%%%
%%%%%%%%%%%%%%%%%%%%%%%%%%%%%%%%%%%%%%%%%%%%%%%%%%%%%%%%%%%%%%%%%%%%%%%%%%%%%%%%%%
\section{Sum rules for the pionic decay amplitudes of $B_2^\ast$ 
up to the order of ${\cal O}(1/m_Q)$ }
\label{sec7}

In the following we shall consider the ${\cal O}(1/m_Q)$ correction to the pionic 
decay amplitude of $B_2^\ast$ meson. There are two decay channels. 
The decay amplitudes are:
\begin{eqnarray}
\label{c1}
 M(B_2^*\to B\pi)&=&I\;\eta_{\mu\nu}q_t^{\mu}q_t^{\nu}
 [g^a(B_2^*,B)+g_1^a(B_2^*,B)]\;,\\
 \label{c2}
 M(B_2^*\to B^*\pi)&=&I\;i\varepsilon_{\alpha\beta\sigma\rho}\;\epsilon^{*\alpha}v^{\beta}
 \eta^{\sigma\mu}q_t^{\rho}q_{t\mu}[g^b(B_2^*,B^*)+g_1^b(B_2^*,B^*)]\;,
\end{eqnarray} 
where $\eta_{\mu\nu}$ is the polarization tensor of $B_2^\ast$ meson, 
$g^a$, $g^b$ are the coupling constants at the leading order of $1/m_Q$ 
and $g_1^a$, $g_1^b$ is the ${\cal O}(1/ m_Q)$ correction to $g^a$, $g^b$.
It can be shown \cite{falk} by combining heavy quark symmetry
and chiral symmetry that at the leading order of $1/m_Q$ 
the coupling constants in (\ref{coup1}), (\ref{c1}) and (\ref{c2}) satisfy 
\begin{eqnarray}
\label{ggg}
g^a(B_2^*,B)=g^b(B_2^*,B^*)=\sqrt{\frac{2}{3}}\;g(B_1,B^*) \; .
\end{eqnarray} 
Angular momentum conservation requires that both operators ${\cal K}$ and ${\cal S}$ 
contribute to the D-wave amplitude only, which is confirmed through explicit 
calculation. Moreover we do not consider mixing effects for $B_2^\ast$ meson, 
which is expected to be small.

Since the detailed calculation is very similar to that for $B_1$ meson, we present 
the final results only:
\begin{eqnarray}\label{final-e}\nonumber
&[(r_{0,-,{1\over 2}} +r_{2,+,{3\over 2}})g^a + g^a_1- 
{\delta m_{1\over 2} +\delta m_{3\over 2} \over T}g^a ] 
 f_{-,{1\over 2} } f_{+, {3\over 2} }\\ \nonumber
&= e^{{ {\bar \Lambda}_{-,{1\over 2} } +{\bar \Lambda}_{+,{3\over 2} } \over T }} \{
{F_\pi\over m_Q}
\int {\cal D} \alpha_i {\alpha_2\over \alpha_3} 
[ \varphi_\bot (\alpha_i) 
-{1\over 2}\tilde\varphi_{\|} (\alpha_i) ] \\ 
&-{\sqrt{2}\over 4}{f_{3\pi}\over m_Q}
\int {\cal D} \alpha_i {1\over \alpha_3} 
 \frac{d}{d\alpha_2}\left( \alpha_2\varphi_{3\pi} (\alpha_i)\right) 
  Tf_0({\omega_c\over T})
 +{1\over 8}{F_{\pi}\over m_Q}K(u_0,T)
\}\;, 
\end{eqnarray}
\begin{eqnarray}\label{final-f}\nonumber
&[(r_{1,-,{1\over 2}} +r_{2,+,{3\over 2}})g^b + g^b_1- 
{\delta m_{1\over 2} +\delta m_{3\over 2} \over T}g^b ] 
 f_{-,{1\over 2} } f_{+, {3\over 2} }\\ 
&= {1\over 2}{F_\pi\over m_Q}
e^{{ {\bar \Lambda}_{-,{1\over 2} } +{\bar \Lambda}_{+,{3\over 2} } \over T }} \{
\int {\cal D} \alpha_i {\alpha_2\over \alpha_3} 
\tilde\varphi_{\|} (\alpha_i) +{1\over 4}K(u_0,T)\}\;, 
\end{eqnarray}
where the function $K(u_0,T)$ is defined in (\ref{KKK}). 

Note that the corrections to the leading order coupling 
constants from the kinetic operator ${\cal K}$ satisfy the same relation 
as (\ref{ggg}), which is an expected result since ${\cal K}$ does not 
depend on the spin.

%%%%%%%%%%%%%%%%%%%%%%%%%%%%%%%%%%%%%%%%%%%%%%%%%%%%%%%%%%%%%%%%%%%%%%%%%%%%%%%%%%%%%
%%%%%%%%%%%%%%%%%%%%%%%%%%%%%%%%%%%%%%%%%%%%%%%%%%%%%%%%%%%%%%%%%%%%%%%%%%%%%%%%%%%%%
\section{Numerical analysis of the sum rules}
\label{sec8}
In the numerical analysis we shall take $T_1=T_2=2T$ and $u_0=1/2$ as
done in \cite{bely95} and {\cite{col97}. Although the initial and final
heavy meson states are different in their masses, this seems to be a
reasonable approximation in view of the large value of $T_1$ and $T_2$
used below.

Note that our sum rules depend on the pion wave functions and 
the integrals or derivatives of them 
at the point $u_0={1\over 2}$. Since $\delta^2$ is numerically small, 
the uncertanty due to the integral term $G_2(u_0)$ is insignificant.
The values of the wave functions at $u_0$ 
have been determined quite well and their dependence on the renormalization scale 
is very weak \cite{bely95}. The first derivatives 
at $u_0$ are zero. The second derivatives 
are very sensitive to the detailed shape of the wave functions. 
Usually only the lowest few moments of the wave functions 
can be determined by QCD sum rules. 
There are many wave functions satisfying these constraints from moments.
With the form of wave functions in \cite{bely95}, the final sum rules are  
sensitive to the renormalization scale and the coefficients $a_i$ in 
$\varphi_\pi (u)$ due to its large value.
This fact indicates
that the not-well-known nonperturbative contribution from 
condensates in the QCD vacuum and the even higher twist wave functions might be 
important. In order to estimate the corrections from the above sources
we present the detailed expressions of the pion wave functions and their 
second derivatives in order to show the dependence on the parameters 
$\omega_{i,j}$ in them.

There are many discussions about $\varphi_\pi (u)$ in literature 
\cite{cz,dziem,bely95,ht,halperin,rad}. The model wave function 
for $\varphi_\pi (u)$ based on the QCD sum rule approach 
was given in \cite{bely95} as:
\begin{equation}\label{wf-0}
\varphi_\pi (u,\mu) =6u{\bar u}\left(
1+a_2(\mu){3\over 2}[5(u-{\bar u})^2-1]+a_4(\mu) {15\over 8} 
[21(u-{\bar u})^4-14(u-{\bar u})^2+1]\right).
\end{equation}
Yet the authors pointed out that the oscillations of (\ref{wf-0}) 
around $u_0={1\over 2}$ is unphysical,
which is due to the truncation of the series and keeping only the first a few 
terms when $\varphi_\pi (u)$ is expanded over Gegenbauer polynominals. In 
\cite{halperin} it was stressed that: (1) the expansion over Gegenbauer polynominals 
converges very slowly as can be seen from the large value of $a_2$ and $a_4$; 
(2) any oscillating wave function is not physical, since no detached scale is seen
to govern such oscillations. Recently Mikhailov and Radyushkin \cite{rad} 
reanalysed the QCD sum rules for $\varphi_\pi (u)$ taking into 
account the nonlocality of the condensates. They suggested  the following 
wave function:
\begin{equation}\label{wf-a}
\varphi_\pi (u) ={8\over \pi}\sqrt{u(1-u)}\;,
\end{equation}
which is close to the asymptotic form and the suggested pion wave 
function in \cite{halperin}:
\begin{equation}\label{wf-b}
\varphi_\pi (u) =N exp\left( -{m^2\over 8\beta^2 u (1-u)} \right)
[\mu^2+\mu {\tilde \mu}+({\tilde\mu}^2-2)u(1-u)]
\;,
\end{equation}
where $N=4.53$, $\mu={m\over 2\beta}$, ${\tilde\mu}={1\over 2\beta}\left(
{1\over 4}m_\pi +{3\over 4}m_\rho\right)={{\tilde m}\over 2\beta}$ with 
$m=330$MeV, ${\tilde m}=620$MeV, and $\beta=320$MeV.
The Brodsky-Huang-Lepage wave function is also close to the 
asymptotic form except for the different behavior at end points \cite{ht}:
\begin{equation}\label{wf-c}
\varphi_\pi (u) =9.05 u(1-u)exp\left( -{M^2\over 8\alpha^2 u (1-u)} \right)
\;,
\end{equation}
where $M=289$MeV and $\alpha =385$MeV. The asymptotic form of $\varphi_\pi (u)$
reads:
\begin{equation}\label{wf-d}
\varphi^{\mbox{asym}}_\pi (u) =6u(1-u)
\;.
\end{equation}

For the other pion wave functions we use the results given in \cite{bely95}. 
The detailed expressions of the pion wave functions relevant in our calculation are:
\begin{eqnarray}\label{fun-a}\nonumber
\varphi_{3\pi}(\alpha_i) =360\alpha_1\alpha_2\alpha_3^2 [1 +\omega_{1,0} {1\over 2}
(7\alpha_3-3) +\omega_{2,0} (2-4\alpha_1\alpha_2-8\alpha_3+8\alpha_3^2) &\\ 
+\omega_{1,1} (3\alpha_1 \alpha_2 -2\alpha_3 +3\alpha_3^2) ] & \; ,
\end{eqnarray}
\begin{equation}\label{wf-2}
\varphi_P (u,\mu) =
1+B_2(\mu){1\over 2}[3(u-{\bar u})^2-1]+B_4(\mu) {1\over 8} 
[35(u-{\bar u})^4-30(u-{\bar u})^2+3],
\end{equation}
\begin{equation}\label{wf-3}
\varphi_\sigma (u,\mu) =6u{\bar u}\left(
1+C_2(\mu){3\over 2}[5(u-{\bar u})^2-1]+C_4(\mu) {15\over 8} 
[21(u-{\bar u})^4-14(u-{\bar u})^2+1]\right),
\end{equation}
\begin{equation}\label{fun-b}
\varphi_\bot (\alpha_i) =30\delta^2 (\alpha_1 -\alpha_2) \alpha_3^2 [{1\over 3}
 +2\epsilon(1-2\alpha_3)] \; ,
 \end{equation}
\begin{equation}\label{fun-c}
\tilde\varphi_{\|} (\alpha_i) =-120\delta^2 \alpha_1 \alpha_2 \alpha_3 [{1\over 3}
 +\epsilon(1-2\alpha_3)] \; ,
 \end{equation}
\begin{eqnarray}\label{wf-4}\nonumber
g_1(u)={5\over 2}\delta^2{\bar u}^2u^2+{1\over 2}\epsilon \delta^2 
[{\bar u}u(2+13u{\bar u})+10u^3\ln u (2-3u+{6\over 5}{\bar u}^2) &\\
+10{\bar u}^3\ln {\bar u}(2-3{\bar u}+{6\over 5}{\bar u}^2)&\;,
\end{eqnarray}
\begin{equation}\label{wf-5}
g_2(u)={10\over 3}\delta^2 {\bar u}u (u-{\bar u}) \;,
\end{equation}
\begin{equation}\label{wf-6}
G_2(u)=-{10\over 3}\delta^2 u^3({2\over 5}u^2-{3\over 4}u+{1\over 3}) \;,
\end{equation}
where ${\bar u}=1-u$, $B_2=30R$, $B_4={3\over 2}R(4\omega_{2,0}-\omega_{1,1}-
2\omega_{1,0})$, $C_2=R(5-{1\over 2}\omega_{1,0})$, 
$C_4={1\over 10}R(4\omega_{2,0}-\omega_{1,1})$ and $R={f_{3\pi}\over \mu_\pi f_\pi}$.
The coefficients $f_{3\pi}$, $\omega_{i,k}$ \cite{f-omega}, $\delta^2$ \cite{delta2} 
and $\epsilon$ \cite{epsilon} have been determined from QCD sum rules.

We shall take the scale $\mu ={3\over 4}\omega_c =2.4$GeV, 
at which the various parameters are:
$a_2 =0.35$, $a_4=0.18$,
$\omega_{1,0}=-2.18$,  $\omega_{2,0}=8.12$,  $\omega_{1,1}=-2.59$, 
$f_{3\pi}=0.0026$GeV$^2$, $\mu_\pi =2.02$GeV, 
$\delta^2 =0.17$GeV$^2$, $\epsilon =0.36$. 

\begin{equation}\label{wf-11}
\varphi_\pi (u_0) ={3\over 2}(1-{3\over 2}a_2 +{15\over 8}a_4),
\end{equation}
\begin{equation}\label{fun-aa}
\varphi_P (u_0) =1-{1\over 2}B_2 +{3\over 8}B_4,
\end{equation}
\begin{equation}\label{fun-bb}
\varphi_\sigma (u_0) ={3\over 2}(1-{3\over 2}C_2 +{15\over 8}C_4),
\end{equation}
\begin{equation}\label{wf-33}
g_1(u_0)=({5\over 32}-0.037\epsilon )\delta^2,
\end{equation}
\begin{equation}\label{wf-44}
G_2(u_0)=-{7\over 288}\delta^2,
\end{equation}
\begin{equation}\label{wf-111}
\varphi^{\prime\prime}_\pi (u_0) =-12(1-9a_2 +{225\over 8}a_4),
\end{equation}
\begin{equation}\label{fun-bbb}
\varphi^{\prime\prime}_\sigma (u_0) =-12(1-9C_2 +{225\over 8}C_4),
\end{equation}
\begin{equation}\label{wf-333}
g^{\prime\prime}_1(u_0)=-(2.5+7.4\epsilon )\delta^2,
\end{equation}
\begin{equation}\label{wf-444}
g^\prime_2(u_0)={5\over 3}\delta^2,
\end{equation}
\begin{equation}\label{wf-66}
\int {\cal D} \alpha_i {\alpha_2\over \alpha_3} \varphi_\bot (\alpha_i) 
=-{\delta^2\over 12}(1+2\epsilon )\; ,
\end{equation}
\begin{equation}\label{wf-77}
\int {\cal D} \alpha_i {\alpha_2\over \alpha_3}\tilde\varphi_{\|} (\alpha_i)
=-({2\over 3} +\epsilon )\delta^2 \; ,
\end{equation}
\begin{equation}\label{wf-88}
\int {\cal D} \alpha_i {1\over \alpha_3} 
 \frac{d}{d\alpha_2}[\alpha_2\varphi_{3\pi} (\alpha_i)]
 =6-2\omega_{1,0}-{4\over 7}\omega_{2,0}+{8\over 7}\omega_{1,1} \; ,
\end{equation} 
where the parameters $a_i$, $B_i$, $C_i$, $\omega_{i,j}$ and $\epsilon$ 
indicate the deviation from the asymtotic form of pion wave functions.

At $u_0={1\over2}$ the values of the functions appearing 
in (\ref{final-a})-(\ref{final-e}) are: $\varphi_\pi(u_0)=1.22$, 
$1.273$, $1.25$, $1.71$, $1.5$ for the wave functions 
(\ref{wf-0})-(\ref{wf-d}) resepctively,
$\varphi_P(u_0)=1.07$, $\varphi_\sigma(u_0)=1.55$,  
$g_1(u_0)=0.022 $GeV$^2$, $g_2 (u_0) =0$ and 
$G_2(u_0)=-4.1\times 10^{-3}$GeV$^2$. Their first derivatives satisfy:
$\varphi^\prime_\pi(u_0)=\varphi^\prime_P(u_0) =\varphi^\prime_\sigma(u_0)=
g^\prime_1(u_0)=G_2^\prime(u_0)=0$. And their second derivatives are:
$\varphi^{\prime\prime}_\pi(u_0)=-35$, $-5.09$, $-0.02$, $-17.50$, $-12$
for (\ref{wf-0})-(\ref{wf-d}) resepctively,
$\varphi^{\prime\prime}_\sigma(u_0)=-17$,  
$g^{\prime\prime}_1(u_0)=-0.88 $GeV$^2$, $g^\prime_2 (u_0) =0.28$GeV.

Note the theoretical side of 
the sum rules (\ref{final-b}), (\ref{final-e}) and (\ref{final-f}) 
depends on $\varphi_\pi (u)$ only through the combination 
$[u^3\varphi_\pi (u)]^{\prime\prime}_{u_0} =3\varphi_\pi (u_0)
+{1\over 8}\varphi^{\prime\prime}_\pi(u_0)$ in $K (u_0, T)$. 
Its value is $3.18$, 
$3.75$, $2.93$ and $3$ for (\ref{wf-a})-(\ref{wf-d}) resepctively, 
which is relatively close to each other. We shall present numerical results
using the wave function (\ref{wf-c}).

The dependence on the Borel parameter $T$ of $f_{-,{1\over 2}} f_{+,{3\over 2}}g$, 
$f_{-,{1\over 2}} f_{+,{1\over 2}}g'$, 
$m_Qf_{-,{1\over 2}} f_{+,{3\over 2}}g_{1s}$, 
$m_Qf_{-,{1\over 2}} f_{+,{3\over 2}}\left((r_{1,-,{1\over 2}} +r_{1,+,{3\over 2}})g + g_{1d}\right)$, 
$m_Qf_{-,{1\over 2}} f_{+,{3\over 2}}\left((r_{0,-,{1\over 2}} +r_{2,+,{3\over 2}})g^a + g^a_1\right)$,  
and  \\
$m_Qf_{-,{1\over 2}} f_{+,{3\over 2}}\left((r_{1,-,{1\over 2}} +r_{2,+,{3\over 2}})g^b + g^b_1\right)$ 
are shown in Fig. 4-10 with $\omega_c=3.2, 3.0, 2.8$ GeV 
and the pion wave function (\ref{wf-c}). 
Requiring that the higher twist contribution is less than $30\%$ of the whole 
sum rule, we get the lower limit for the Borel parameter $T$, $T\ge 1.0$GeV. 
Requiring that the continuum contribution is less than $40 \%$ of the whole sum rule,
we get the upper limit, $T\le 2.5$GeV.  

Our results read:
\begin{equation}\label{num-a}
f_{-,{1\over 2}} f_{+,{3\over 2}}g=-(0.23\pm 0.02\pm 0.01)~~\mbox{GeV}^3\;,
\end{equation}
\begin{equation}\label{num-b}
f_{-,{1\over 2}} f'_{+,{1\over 2}}g^\prime =-(0.41\pm 0.03\pm 0.04)~~\mbox{GeV}^5 \;
\end{equation}
with the current (\ref{curr4}), or
\begin{equation}\label{num-bb1}
f_{-,{1\over 2}} f_{+,{1\over 2}}g^\prime =(0.41\pm 0.03\pm 0.01)~~\mbox{GeV}^4 \;
\end{equation}
with the current (\ref{curr2}),
\begin{equation}\label{num-c}
f_{-,{1\over 2}} f_{+,{3\over 2}}g_{1s}= 
{0.038\pm 0.005\pm 0.006\over m_Q}~~\mbox{GeV}^5\;
\end{equation}
\begin{equation}\label{num-d}
f_{-,{1\over 2}} f_{+,{3\over 2}}[(r_{1,-,{1\over 2}} 
+r_{1,+,{3\over 2}})g + g_{1d}]={0.21\pm 0.04 \pm 0.02 \over m_Q}~~\mbox{GeV}^3
\; ,
\end{equation}
\begin{equation}\label{num-e}
f_{-,{1\over 2}} f_{+,{3\over 2}}[(r_{0,-,{1\over 2}} +r_{2,+,{3\over 2}})g^a 
+ g^a_1]={0.17\pm 0.01\pm 0.02 \over m_Q}~~\mbox{GeV}^3
\; ,
\end{equation}
\begin{equation}\label{num-f}
f_{-,{1\over 2}} f_{+,{3\over 2}}[(r_{1,-,{1\over 2}} +r_{2,+,{3\over 2}})g^b +
 g^b_1]={0.17\pm 0.01 \pm 0.02\over m_Q}~~\mbox{GeV}^3
\; .
\end{equation}
The central values conrespond to $T=1.3$GeV and $\omega_c=3.0$GeV. The first 
and second error refers to  
the variation with $T$ and $\omega_c$ respectively. The inherent uncertainties 
due to the method of QCD sum rules and the pion wave functions are not 
included here.

With the central values of f's in (\ref{ff1})-(\ref{ff3}) and 
the values of r's in section \ref{sec3}, we arrive at:
\begin{equation}\label{nu-a}
g=-(4.83\pm 0.4\pm 0.2)~~\mbox{GeV}^{-2}\;,
\end{equation}
\begin{equation}\label{nu-b}
g^\prime =-(4.43\pm 0.33\pm 0.44) \; 
\end{equation}
with the current (\ref{curr4}), or
\begin{equation}\label{nu-b1}
g^\prime =-(4.1\pm 0.3\pm 0.3) \; 
\end{equation}
with the current (\ref{curr2}),
\begin{equation}\label{nu-c}
g_{1s}={0.77\pm 0.19\pm 0.19\over m_Q}\;,
\end{equation}
\begin{equation}\label{nu-d}
g_{1d}=-{2.1\pm 1.0 \over m_Q}~~\mbox{GeV}^{-2}
\; ,
\end{equation}
\begin{equation}\label{nu-e}
g^a_1=-{1.6\pm 0.8 \over m_Q}~~\mbox{GeV}^{-2}
\; ,
\end{equation}
\begin{equation}\label{nu-f}
g^b_1=-{1.6\pm 0.8 \over m_Q}~~\mbox{GeV}^{-2}
\; .
\end{equation}
We have not included the errors due to the uncertainty of f's in 
(\ref{nu-a})-(\ref{nu-f}), which is about $20\%$.

Therefore, with the pion wave function $\varphi_\pi (u)$ in (\ref{wf-c}) 
the D-wave coupling constants are
$g_d(B_1,B^\ast)=-(4.83 +{2.1\over m_Q})=-5.27$GeV$^{-2}$, 
$g_d(B_2^\ast,B)=-(3.94 +{1.6\over m_Q})=-4.27$GeV$^{-2}$, 
$g_d(B_2^\ast,B^\ast)=-(3.94 +{1.91\over m_Q})=-4.34$GeV$^{-2}$ with $m_b=4.8$GeV. 
Strong cancellation ocurrs for ${\cal O}(1/m_Q)$
corrections to the coupling constants from the operator 
${\cal K}$. At the phenomenological side the factor 
$(r_{1,-,{1\over 2}}+r_{1,+,{3\over 2}})$ is somewhat large. Luckily the 
${\cal O}(1/m_Q)$ correction on the theoretical side has the same 
sign as $(r_{1,-,{1\over 2}}+r_{1,+,{3\over 2}})g$. Moreover their
magnitude is close to each other. Their contribution cancels 
largely. As the result the ${\cal O}(1/m_Q)$ correction is about
of $9\%$ for the $B_1$ and $8\%$ for $B^\ast_2$ meson respectively.
As mentioned above,
since $(\ref{final-b})$, $(\ref{final-e})$ and $(\ref{final-f})$ depend
on the second derivative of $u^3 \varphi_\pi (u)$ at $u_0 ={1\over 2}$, 
the ${\cal O}(1/m_Q)$ correction to the D-wave coupling constants 
has a weak dependence on the the detailed shape of
$\varphi_\pi (u)$ around $u_0 ={1\over 2}$. 
For example, if we use the pion wave function $\varphi_\pi (u)$ in (\ref{wf-a}),  
the ${\cal O}(1/m_Q)$ correction to the D-wave coupling constants is about
$4\%$ for the $B_1$ and $3\%$ for $B^\ast_2$ meson respectively. 
The ${\cal O}(1/m_Q)$ correction has the same sign as the leading order 
coupling constant $g$ using all the five forms of 
$\varphi_\pi (u)$ in (\ref{wf-0})-(\ref{wf-d}). 
 
%%%%%%%%%%%%%%%%%%%%%%%%%%%%%%%%%%%%%%%%%%%%%%%%%%%%%%%%%%%%%%%%%%%%%%%%%%%%%%%%%%%%%
%%%%%%%%%%%%%%%%%%%%%%%%%%%%%%%%%%%%%%%%%%%%%%%%%%%%%%%%%%%%%%%%%%%%%%%%%%%%%%%%%%%%%
\section{Pionic widths of mesons in the $(1^+,2^+)$ doublet and discussions}
\label{sec9}

From the above results, it is straightforward to derive the
decay widths for $B_1$ and $B_2^*$. 
Summing over charged and neutral pion channels the decay widths are:
\begin{equation}
\label{w-1}
\Gamma(B_1\to B^*\pi)={1\over 12\pi} 
(g^2+2gg_{1d})|\vec q|^5 +{3\over 8\pi} g^2_{1s}|\vec q| \;,
\end{equation}
\begin{equation}
\label{w-2}
\Gamma(B_2^*\to B\pi)={1\over 20\pi}(g_a^2+2g^ag_1^a) |\vec q|^5\;,
\end{equation}
\begin{equation}
\label{w-3}
\Gamma(B_2^*\to B^*\pi)={3\over 40\pi} (g_b^2+2g^bg_1^b)|\vec q|^5\;.
\end{equation}
The last term in (\ref{w-1}) is formally of ${\cal O}(1/m^2_Q)$. However, 
due to the large space phase factor for the S-wave decay it is 
numerically not small. 

The masses of $B_1, B_2^\ast$ are not known experimentally. 
The decay width is sensitive to the $B_1, B_2^\ast$ masses. 
If we use the pion wave function (\ref{wf-c}) in 
\cite{ht}, the decay widths are:
\begin{equation}
\label{ww-11}
\Gamma(B_1\to B^*\pi)=7.0\times \left( {|\vec q|\over 395\mbox{MeV}}\right)^5
+1.2 \times \left( {\beta\over 0.02}\right)^2
\left( {|\vec q|\over 395\mbox{MeV}}\right)
~\mbox{MeV} \;,
\end{equation}
\begin{equation}
\label{ww-22}
\Gamma(B_2^*\to B\pi)=5.4\times \left( {|\vec q|\over 450\mbox{MeV}}\right)^5
~\mbox{MeV}\;,
\end{equation}
\begin{equation}
\label{ww-33}
\Gamma(B_2^*\to B^*\pi)=4.9\times \left( {|\vec q|\over 405\mbox{MeV}}\right)^5
~\mbox{MeV}\; ,
\end{equation}
where the decay momentum $|\vec q|$ is in unit of MeV respectively.

If we assume the heavy quark expansion is also good for the charm system, 
the decay widths of $D_1$ and $D_2^\ast$ are:
\begin{equation}
\label{ww-111}
\Gamma(D_1\to D^*\pi)=6.2+9.7 \times \left( {\beta\over 0.06}\right)^2
~\mbox{MeV} \;,
\end{equation}
\begin{equation}
\label{ww-222}
\Gamma(D_2^*\to D\pi)=13.7 ~~\mbox{MeV}\;,
\end{equation}
\begin{equation}
\label{ww-333}
\Gamma(D_2^*\to D^*\pi)=6.1 ~~\mbox{MeV}\; ,
\end{equation}
where we have used the decay momenta $|\vec q|=$ 355, 503 and 387
$\mbox{Mev}$ respectively for the above three processes listed in \cite{review}.
Experimentally, the total decay widths of $D_1$ and $D_2$ are 
$18.9^{+4.6}_{-3.5}$MeV and $23^{+5}_{-5}$MeV respectively.

In summary, in this work we have improved previous results about the kinetic energy ${\cal K}$ and 
chromomagnetic splitting $\Sigma$ of the excited heavy mesons in HQET. 
We have carried out a systematic expansion up to the order
 ${\cal O}(1/m_Q)$ for the pionic decay amplitutes of $B_1$ and $B_2^*$
in the framework of HQET and light-cone sum rules. The mixing of the
two $1^+$ states has been properly taken into account.
The ${\cal O}(1/m_Q)$  correction to the decay amplitutes of $B_1$ and
$B_2^*$ derived from the light-cone sum rule is small. When applied to 
$D_1$ and $D_2^*$ the central theoretical values are in good agreement
with the experimental data, though the theoretical uncertainty is
sizable for the $D$ system.

\vspace{0.8cm} {\it Acknowledgements:\/} S.Z. was supported by
the National Postdoctoral Science Foundation of China and Y.D. was 
supported by the National Natural Science Foundation of China.

\vspace{1.cm}

\newpage
{\bf Figure Captions}
\vspace{2ex}
\begin{center}
\begin{minipage}{130mm}
{\sf Fig. 1.} \small{The sum rules for $G_K$ as a functions of the Borel parameter
$T$ for different values of the continuum threshold $\omega_c$. From bottom
 to top the curves correspond to $\omega_c=3.2, 3.0, 2.8$ GeV respectively.
 The unit of $T$, $G_K$ is GeV. }
\end{minipage}
\end{center}
\begin{center}
\begin{minipage}{130mm}
{\sf Fig. 2.} \small{The sum rules for $G_\Sigma$
with $\omega_c=3.2, 3.0, 2.8$ GeV with the curves from top to bottom.}
\end{minipage}
\end{center}
\begin{center}
\begin{minipage}{130mm}
{\sf Fig. 3.} \small{The sum rules for 
$G_{{1\over 2},{3\over 2}}/f_{+,{1\over 2}}$ 
with $\omega_c=3.2, 3.0, 2.8$ GeV. }
\end{minipage}
\end{center}
\begin{center}
\begin{minipage}{130mm}
{\sf Fig. 4.} \small{The sum rules for $f_{-,{1\over 2}} f_{+,{3\over 2}}g$ 
with $\omega_c=3.2, 3.0, 2.8$ GeV 
with the pion wave function in (\ref{wf-c}) in \cite{ht}. }
\end{minipage}
\end{center}
\begin{center}
\begin{minipage}{130mm}
{\sf Fig. 5.} \small{The sum rules for $f_{-,{1\over 2}} f_{+,{1\over 2}}g'$ 
with the current (\ref{curr4}) and $\omega_c=3.2, 3.0, 2.8$ GeV.}
\end{minipage}
\end{center}
\begin{center}
\begin{minipage}{130mm}
{\sf Fig. 6.} \small{ The sum rules for $f'_{-,{1\over 2}} f_{+,{1\over 2}}g'$ 
with the current (\ref{curr2}) $\omega_c=3.2, 3.0, 2.8$ GeV.}
\end{minipage}
\end{center}
\begin{center}
\begin{minipage}{130mm}
{\sf Fig. 7.} \small{The sum rules for $m_Qf_{-,{1\over 2}} f_{+,{3\over 2}}g_{1s}$ 
with $\omega_c=3.2, 3.0, 2.8$ GeV.}
\end{minipage}
\end{center}
\begin{center}
\begin{minipage}{130mm}
{\sf Fig. 8.} \small{ The sum rules for  
$m_Qf_{-,{1\over 2}} f_{+,{3\over 2}}\left((r_{1,-,{1\over 2}} +r_{1,+,{3\over 2}})g + g_{1d}\right)$ 
with $\omega_c=3.2, 3.0, 2.8$ GeV.}
\end{minipage}
\end{center}
\begin{center}
\begin{minipage}{130mm}
{\sf Fig. 9.} \small{ The sum rules for  
$m_Qf_{-,{1\over 2}} f_{+,{3\over 2}}\left((r_{0,-,{1\over 2}} +r_{2,+,{3\over 2}})g^a + g^a_1\right)$ 
with $\omega_c=3.2, 3.0, 2.8$ GeV.}
\end{minipage}
\end{center}
\begin{center}
\begin{minipage}{130mm}
{\sf Fig. 10.} \small{ 
The sum rules for $m_Qf_{-,{1\over 2}} f_{+,{3\over 2}}\left((r_{1,-,{1\over 2}} +r_{2,+,{3\over 2}})g^b + g^b_1\right)$ 
with $\omega_c=3.2, 3.0, 2.8$ GeV.}
\end{minipage}
\end{center}
 
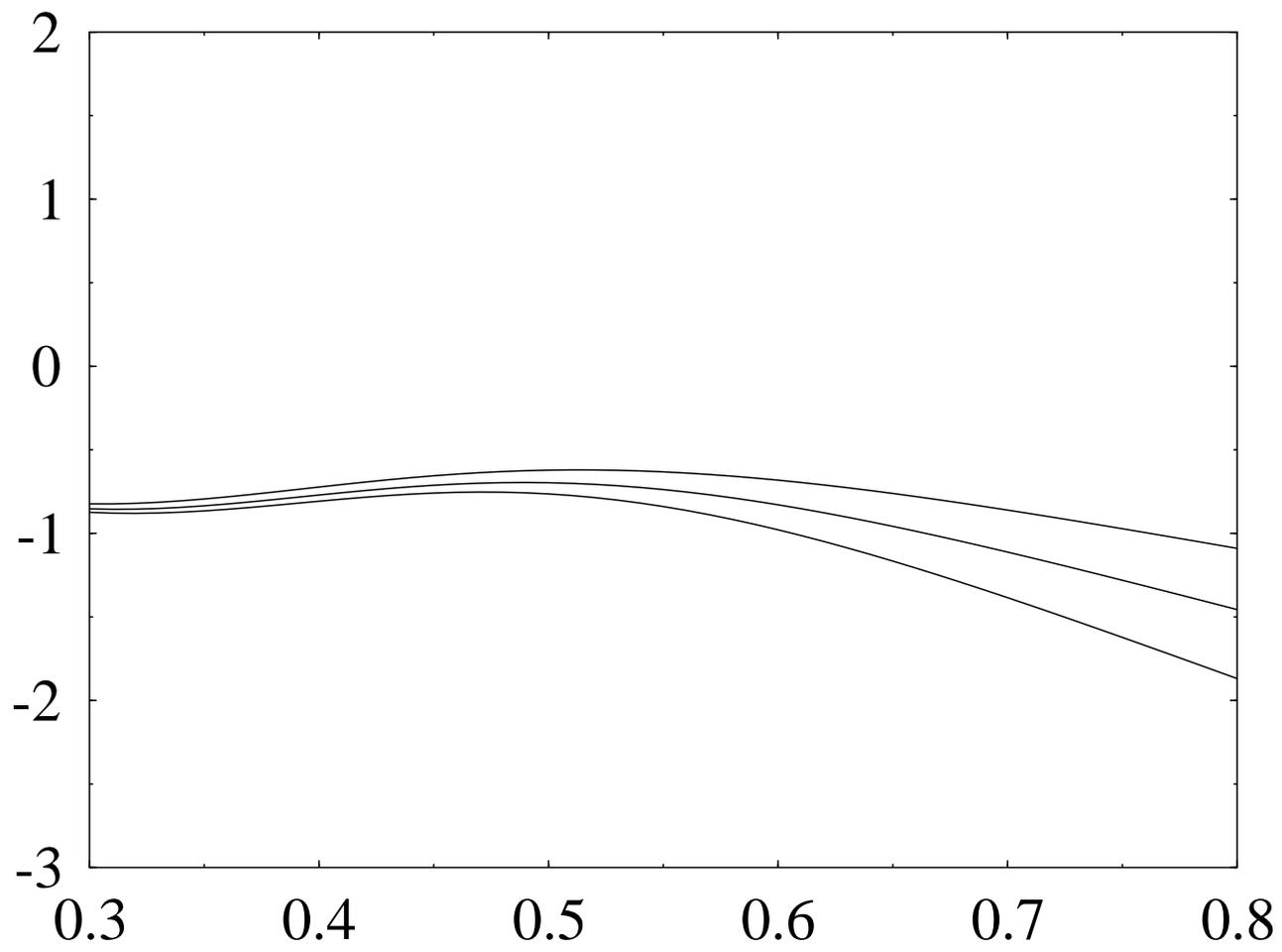
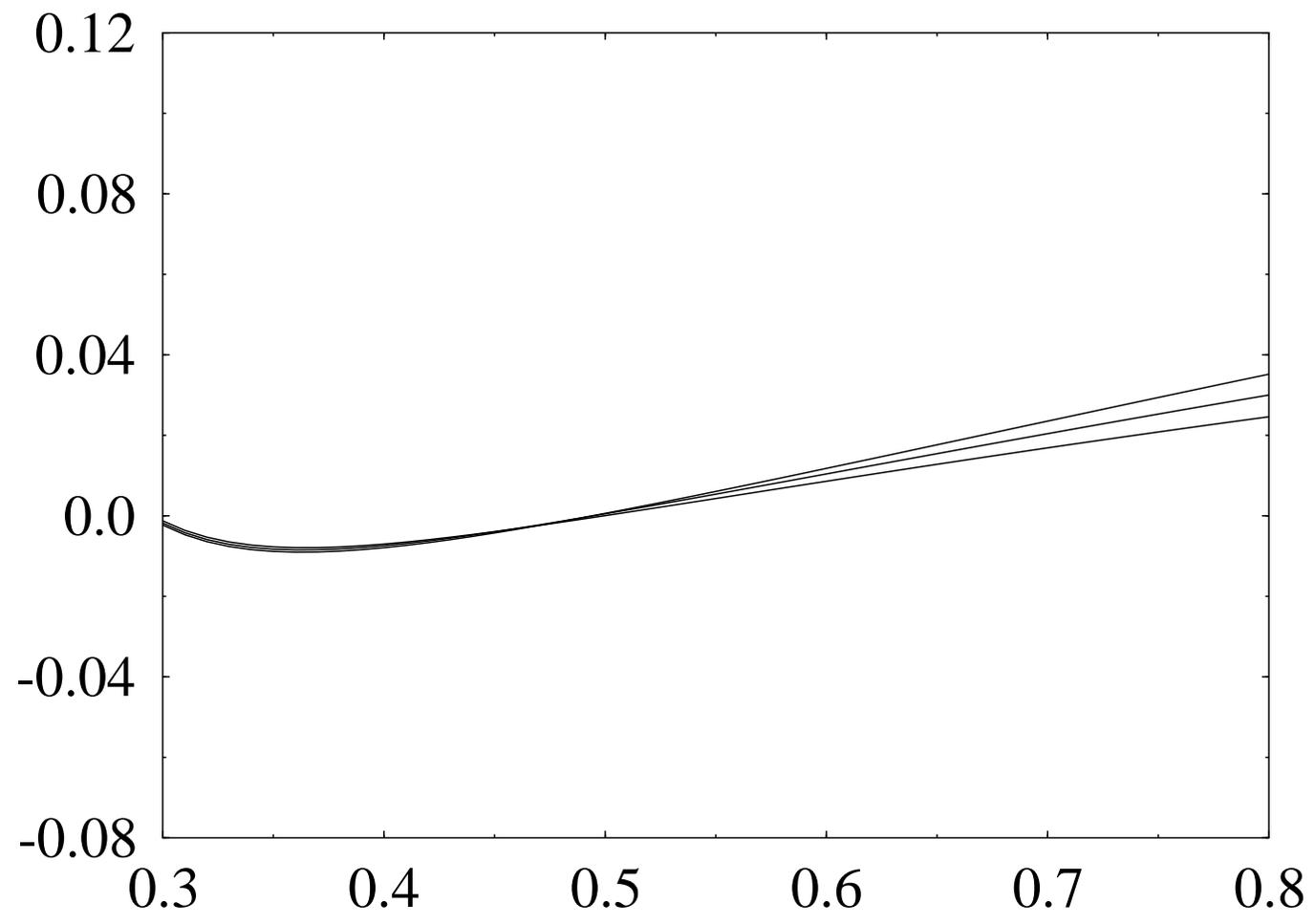
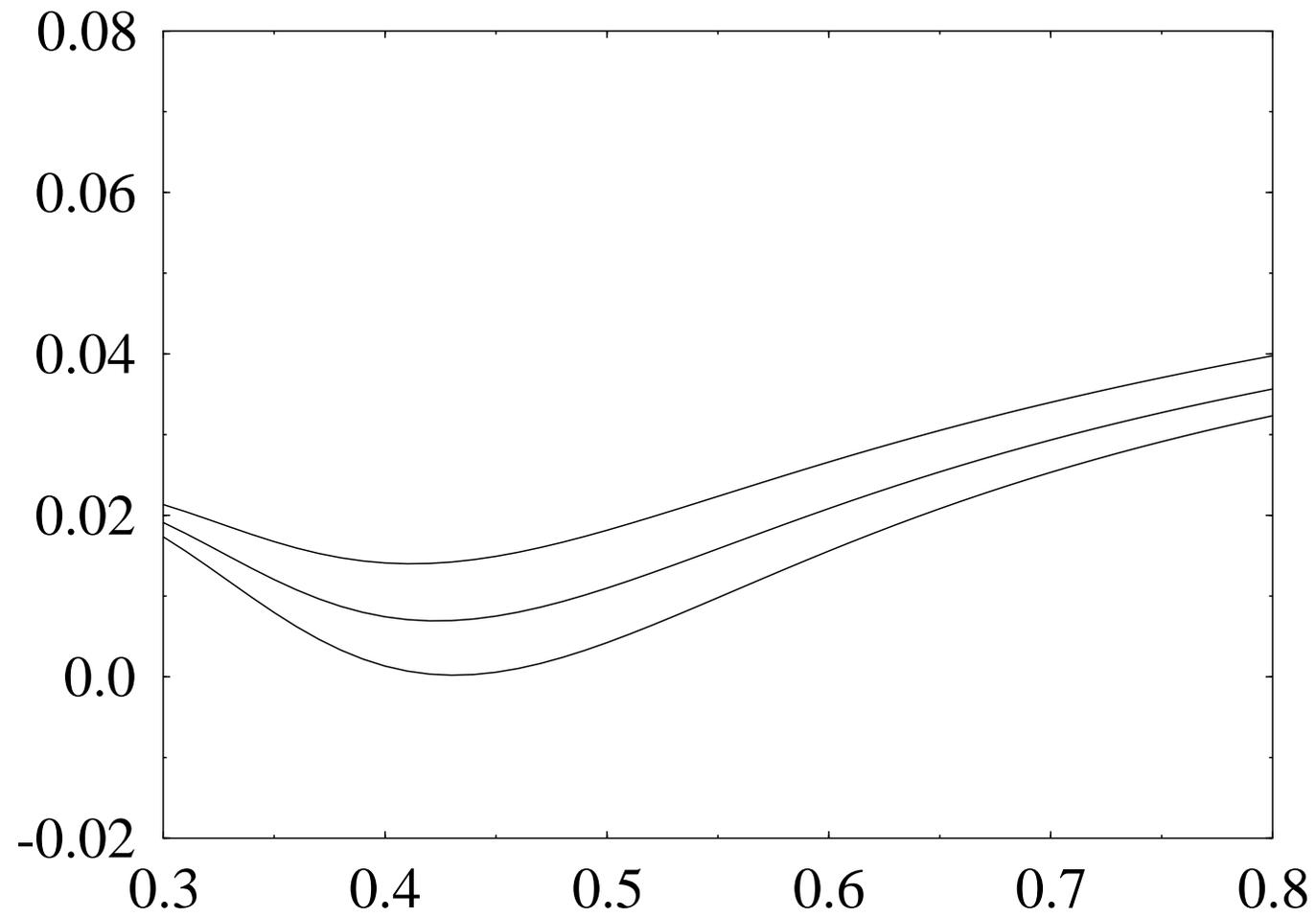
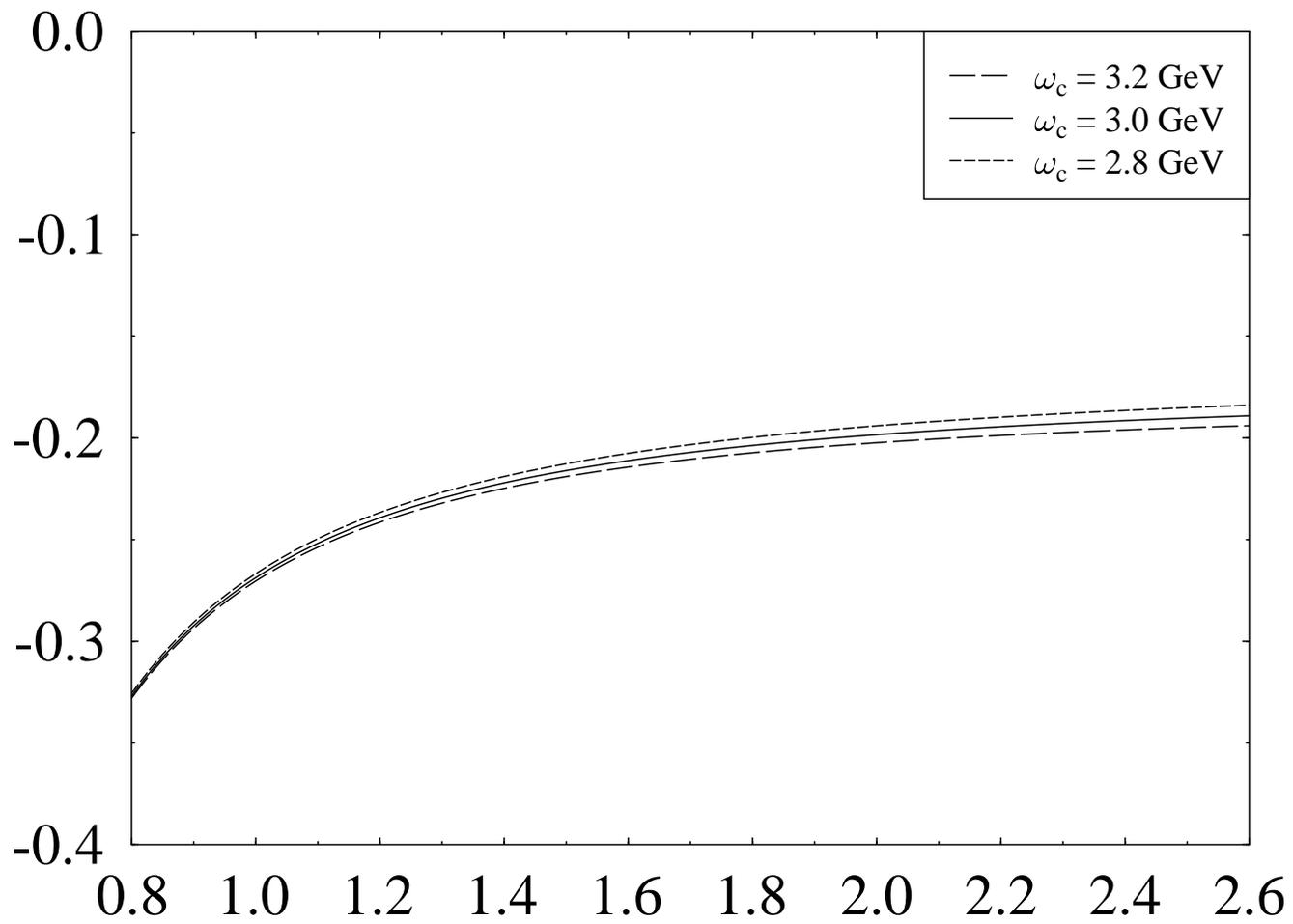
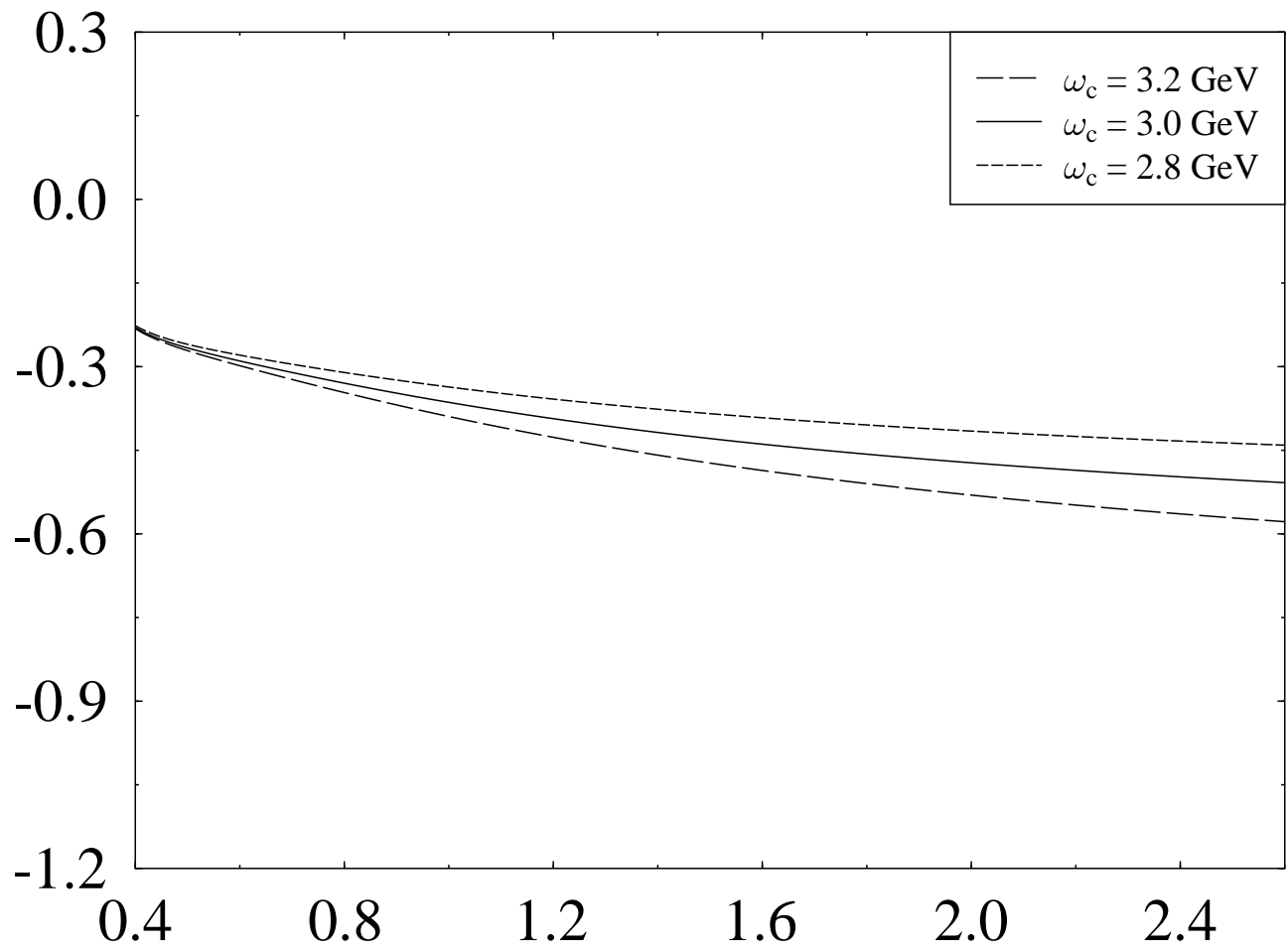
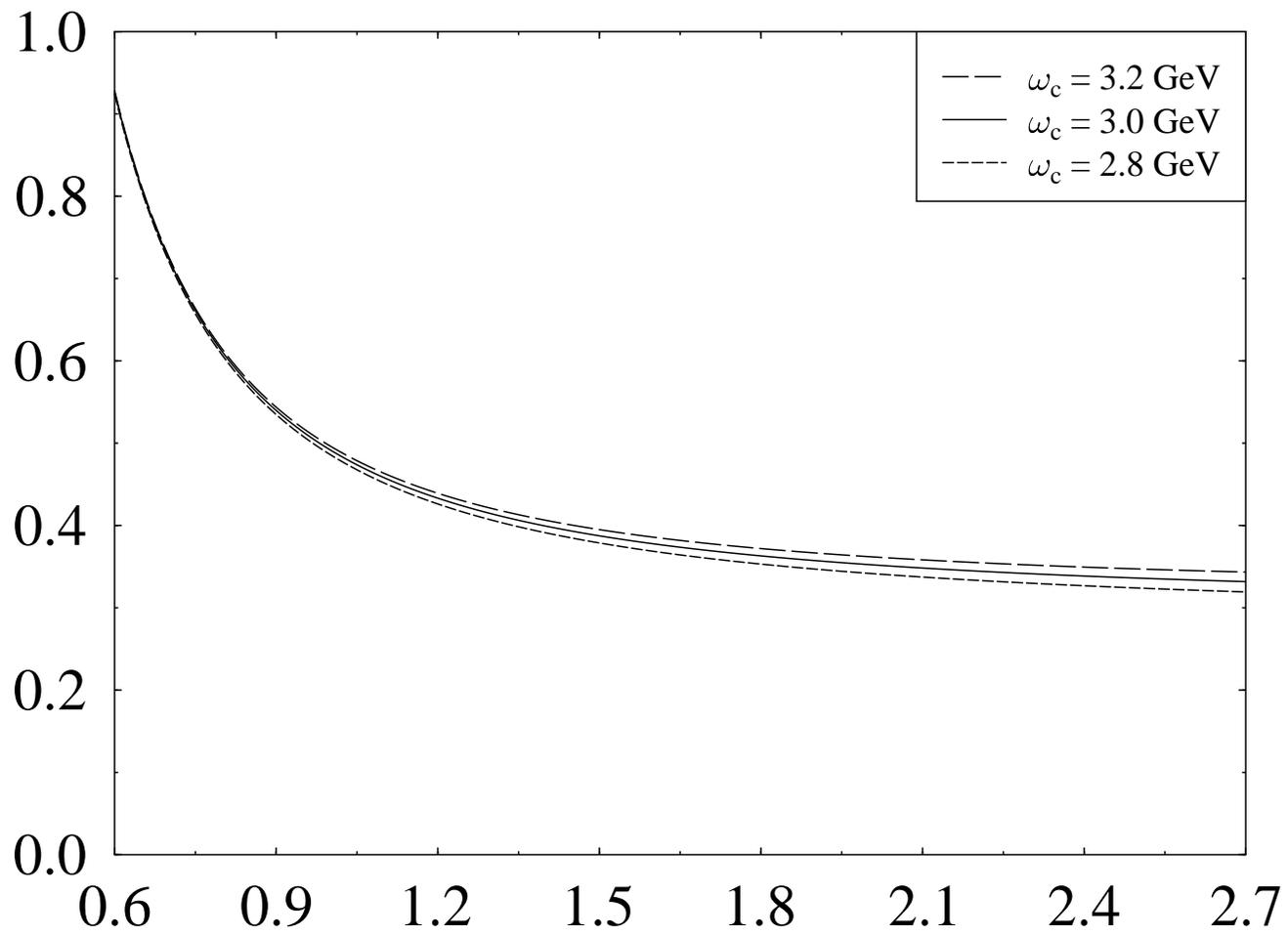
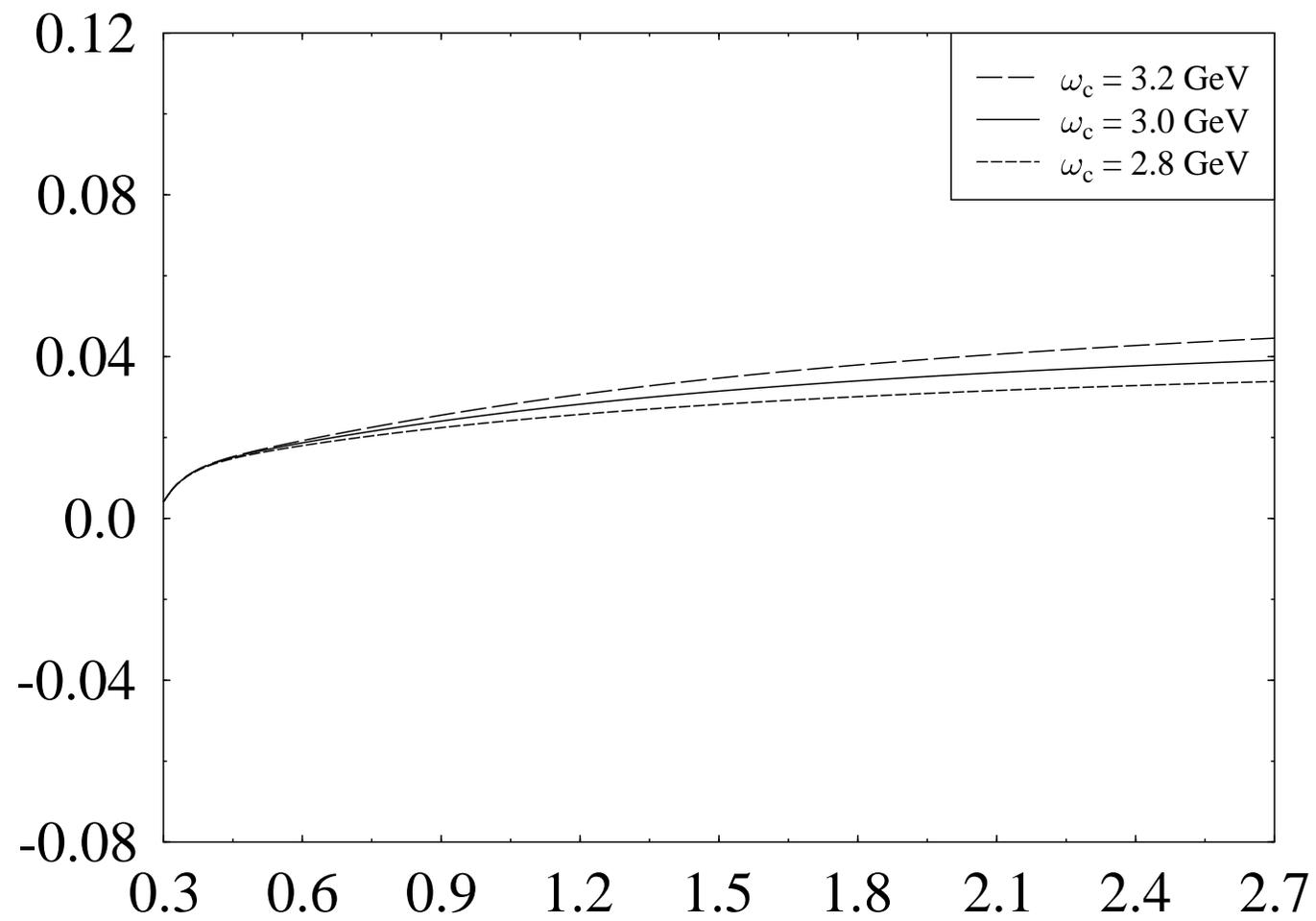
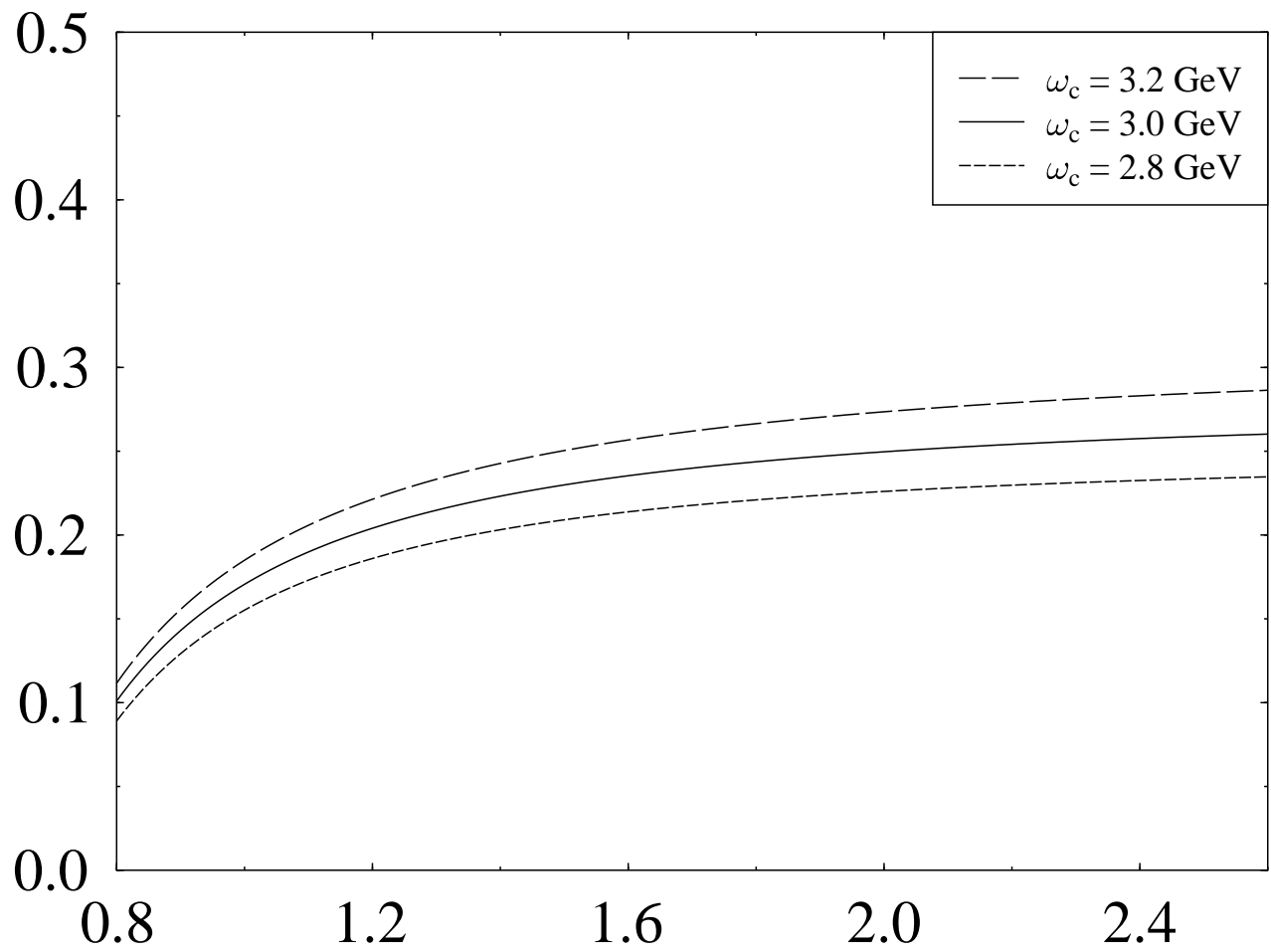
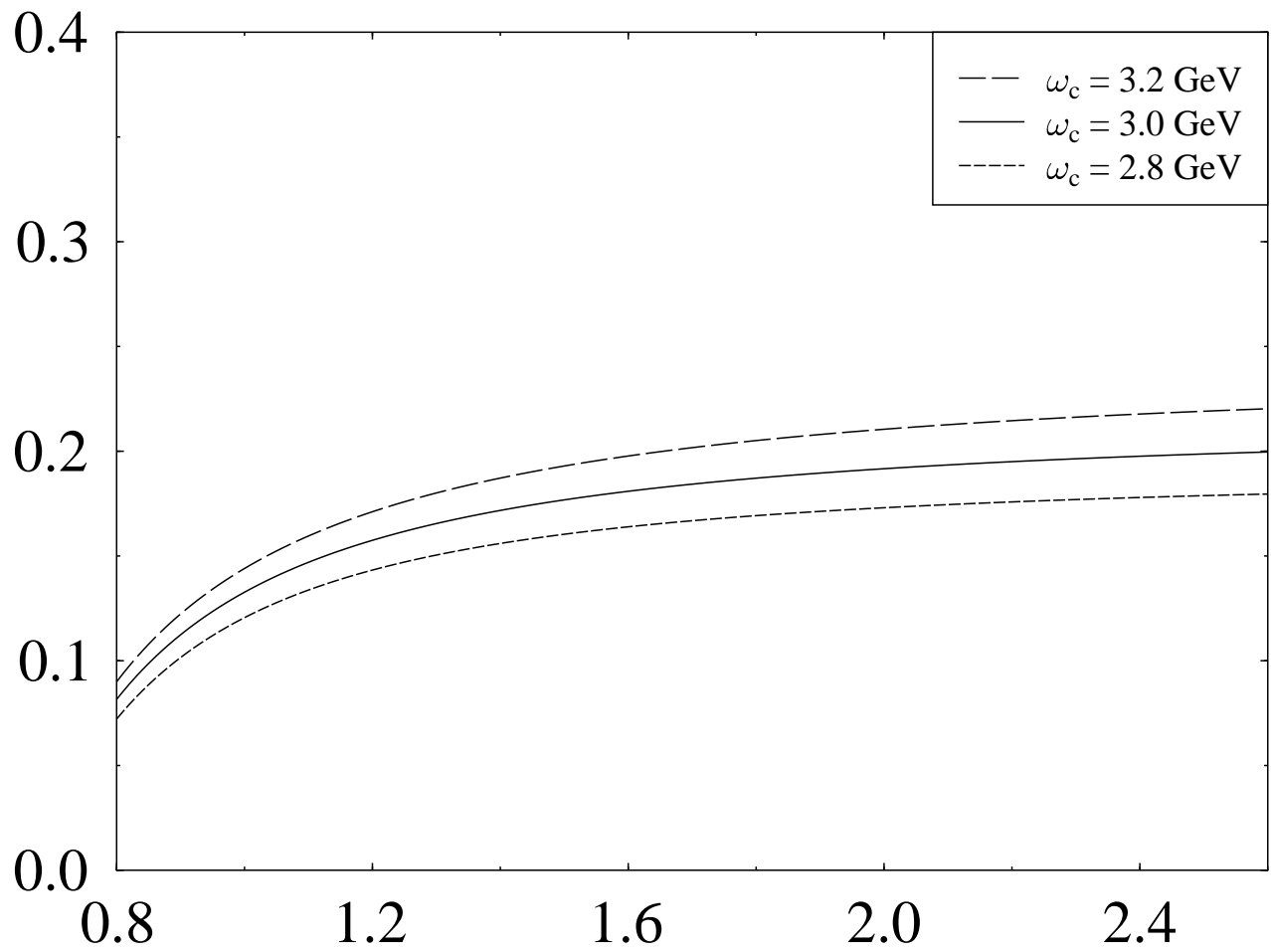
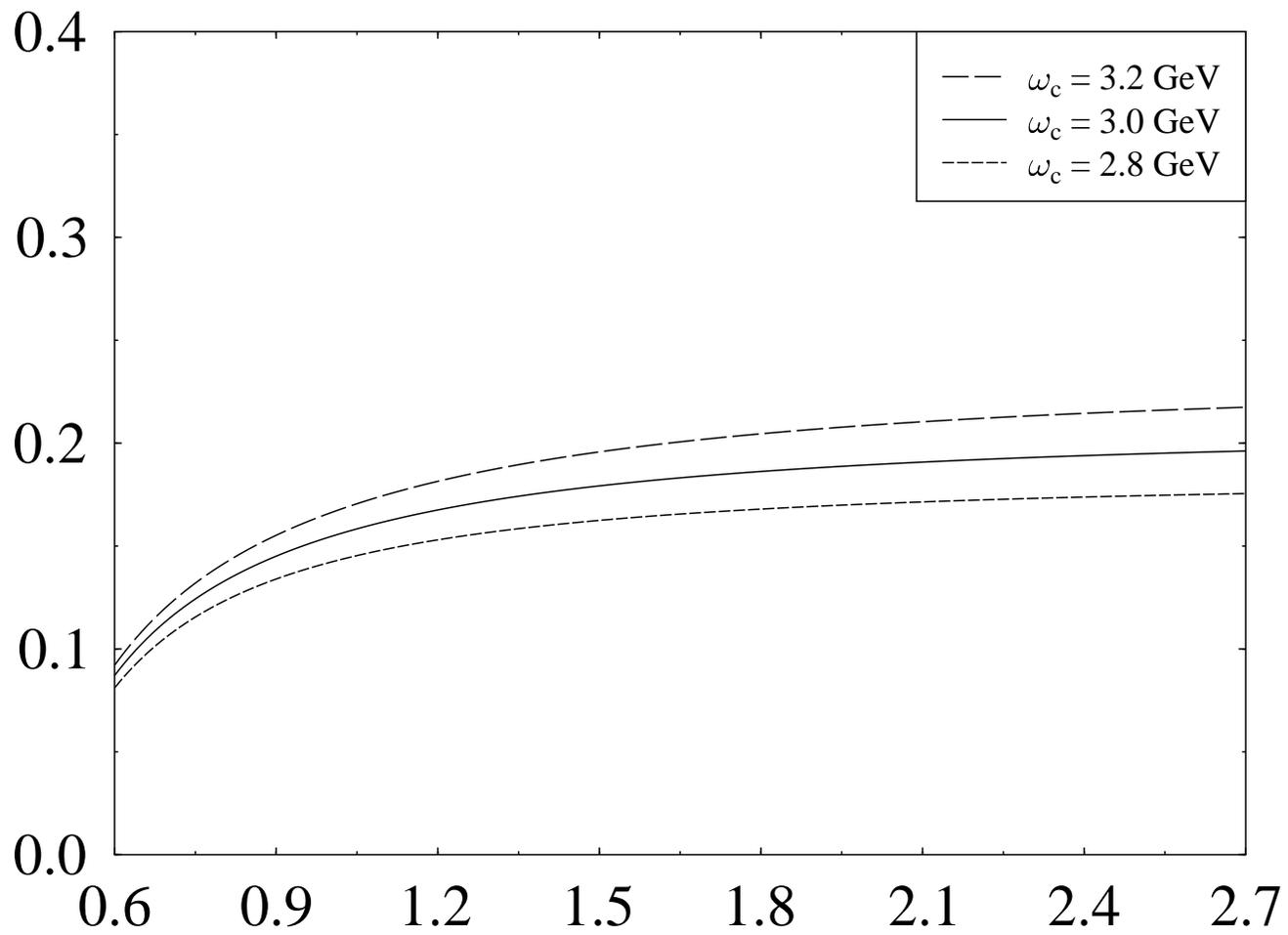
\end{document}